\documentclass[
  twocolumn,
  groupedaddress,
  superscriptaddress,
  amsmath,
  amssymb,
  amsfonts,
  amstext,
  prb,
]{revtex4-2}

\usepackage{dcolumn}
\usepackage{bm}
\usepackage{soul}
\usepackage[flushleft]{threeparttable}
\usepackage{url}
\usepackage{mathrsfs}
\usepackage{braket}
\usepackage{multirow}
\usepackage{array}
\usepackage{booktabs}
\usepackage{float}
\usepackage{lineno}
\usepackage{hyperref}
\usepackage{mathtools}
\usepackage{xcolor}
\usepackage{enumitem}
\usepackage{tikz}
\usepackage{placeins}

\hypersetup{colorlinks=true, urlcolor=blue, citecolor=blue, linkcolor=blue, pdfborder={0 0 0}}

\def\kk{\mathbf{k}}

\begin{document}
\title{
Emergent Anomalous Hall Effect from Surface States in the Altermagnet MnTe Thin Films
}

\author{Yufei Zhao}
\affiliation{Department of Condensed Matter Physics, Weizmann Institute of Science, Rehovot 7610001, Israel}
\affiliation{Department of Physics, The Pennsylvania State University, University Park, Pennsylvania 16802, USA}
\author{Saswata Mandal}
\affiliation{Department of Physics, The Pennsylvania State University, University Park, Pennsylvania 16802, USA}
\author{Chao-Xing Liu}
\affiliation{Department of Physics, The Pennsylvania State University, University Park, Pennsylvania 16802, USA}
\author{Binghai Yan}
\email{binghai.yan@psu.edu}
\affiliation{Department of Condensed Matter Physics, Weizmann Institute of Science, Rehovot 7610001, Israel}
\affiliation{Department of Physics, The Pennsylvania State University, University Park, Pennsylvania 16802, USA}

\date{\today}

\begin{abstract}
Transport measurements on thin films of the prototypical altermagnet MnTe have reported conflicting phenomena of anomalous Hall effects (AHE), including opposite signs and thickness-independent resistivity. Here we resolve these discrepancies by separating bulk and surface contributions to the AHE for different crystal terminations. Using first-principles calculations and symmetry-based effective models, we show that although the bulk hosts a characteristic $g$-wave Fermi surface, surface states within the bulk gap acquire a ferromagnet-like spin polarization and dominate the AHE at experimentally relevant Fermi energies. While the surface magnetization follows the surface spin sublattice, the resulting AHE is uniquely determined by the bulk Néel order for a given termination. Both bulk and surface contributions are closely linked to a small but finite out-of-plane orbital magnetization. Incorporating realistic interfacial chemistry further reveals that a Te capping layer can reverse the surface AHE sign relative to that on an InP substrate. Our results establish a microscopic framework for interpreting and engineering AHE responses in altermagnetic thin films through interface design.
\end{abstract}

\maketitle

\section{Introduction} \label{intro}
Altermagnetism refers to an unconventional collinear antiferromagnetic phase \cite{vsmejkal2020crystal,Yuan2020,ma2021multifunctional,vsmejkal2022beyond,vsmejkal2022emerging,song2025altermagnets,amin2024nanoscale,Mazin2022}. 
Despite vanishing net magnetization, it exhibits nonrelativistic band split and spin current \cite{Wu2007,vsmejkal2022beyond}, and usually shows anomalous Hall effect (AHE) with spin-orbit coupling (SOC) \cite{liu2025different}, which is similar to another known Mn$_3$(Sn/Ge) family of noncollinear antiferromagnets \cite{nakatsuji2015large,nayak2016large,zelezny2017,Zhang2017strong}. In experiments, spin-split bands are typically observed as evidence of altermagnets via angle-resolved photoemission spectroscopy (ARPES) \cite{Lee2024prl,fedchenko2024observation,krempasky2024altermagnetic,zhu2024observation,reimers2024direct,zhang2025crystal}, and the AHE serves as a transport probe of Berry curvature~\cite{gonzalez2023spontaneous,zhao2025nonlinear,subin2025}. 

MnTe emerges as one of the most extensively studied altermagnet materials, owing to its high Néel temperature (310 K) and compatibility with molecular beam epitaxy thin film growth, as a promising candidate for semiconductor spintronics applications. 

MnTe thin films and single crystals were reported to exhibit a variety of AHE behaviors~\cite{gonzalez2023spontaneous, kluczyk2024coexistence, chilcote2024stoichiometry, bey2024unexpected, liu2025strain, smolenski2025strain, zhou2026}. A major puzzle is that the AHE sign sensitively depends on thin film samples grown on different substrates and interface conditions. 
For instance, MnTe films grown on InP substrate typically exhibit a positive anomalous Hall conductivity (AHC) among 50--300 K in early reports \cite{gonzalez2023spontaneous, chilcote2024stoichiometry} and show an AHC sign reversal from positive to negative when cooling down with the critical temperature near 175 K in a more recent work \cite{zhou2026}. In contrast, bulk crystals show similar AHC sign reversal around 230 K where strain was believed to modify AHE ~\cite{liu2025strain,smolenski2025strain}. 
Furthermore, Zhou \textit{et al.}~\cite{zhou2026} found that both the AHE and longitudinal resistance are insensitive to film thickness, raising the question whether surface or interface states govern the magneto-transport properties in these thin films, distinct from bulk crystals.  Zhou \textit{et al.}~\cite{zhou2026} also revealed the existence of surface states at the Fermi energy by ARPES, in addition to bulk valence bands~\cite{Lee2024prl,Osumi2024prb,krempasky2024altermagnetic,hajlaoui2024temperature,chilcote2024stoichiometry} lying 30$\sim$60~meV below the Fermi level.  

Although surface states are known to profoundly influence the electronic and transport properties of low-dimensional systems, their role in altermagnets remains largely unexplored. If surface states dominate the AHE observed in experiments, can we still gain useful insights about the bulk altermagnetism? Therefore, we are motivated to investigate surface states and focus on their interplay with the bulk altermagnetism in MnTe thin films by combining first-principles calculations and effective models in this work. Our results reveal that surface states emerge within the bulk band gap. These states harbor large Berry curvature, thereby generating much stronger AHE than bulk states in the thin film, even when the bulk is $p$-doped {and the film has more than 200 atomic layers}. Because surface band dispersions are sensitive to the surface/interface atomic structures, the AHC sign is surface-termination-dependent, rationalizing different AHC signs in reports \cite{gonzalez2023spontaneous,zhou2026}. For a fixed surface termination, we find that the surface AHE is determined by the bulk altermagnetic order, regardless of which spin sublattice is exposed on the surface. Both first-principles calculations and the effective surface state model analysis reveal that, the out-of-plane component of orbital magnetization---the source of AHE---remains unchanged for opposite spin sublattices on the surface as long as the bulk altermagnetic order remains the same. Our findings establish a unique surface-probe of bulk altermagnetism and pave the way for controlling magnetotransport of altermagnetic devices by surface and interface engineering.

The paper is organized as follows. After introducing the methodology in Section~\ref{method}, we present the surface states at Fermi energy in Section~\ref{spin-FS} and analyze their contribution to the AHE in Section~\ref{AHC_thick}. In Section~\ref{AHC_surfaceM}, we show that the polarity of the surface AHE is dictated by the bulk Néel order and remains robust against variations in the surface magnetization direction for a given termination. In Section~\ref{model}, we introduce a symmetry-based effective model for the surface states to elucidate the mechanism of AHE induced by the out-of-plane orbital moment and orbital texture, instead of surface magnetization. Section~\ref{interface} incorporates realistic interfacial chemistry to examine its impact on the AHE, revealing how different surface terminations (such as Te capping or a MnTe–InP interface) can significantly modify or even invert the AHC. Finally, Section~\ref{discussion} summarizes our findings.

\section{Method} \label{method}
Density-functional theory (DFT) calculations were performed using the Vienna \emph{ab initio} simulation package (\textsc{vasp}) with the projector-augmented-wave (PAW) method \cite{kresse1996efficient,kresse1999ultrasoft}. The exchange-correlation functional was treated within the generalized gradient approximation (GGA) using the Perdew--Burke--Ernzerhof (PBE) parametrization \cite{perdew1996generalized}. To account for electronic correlations in Mn-$3d$ states, we employed the DFT+$U$ method \cite{dudarev1998electron} with on-site Coulomb interaction $U=4.8$~eV and exchange interaction $J=0.8$~eV. Maximally localized Wannier functions for Mn-$s,d$, Te-$p$, In-$s,p$, P-$p$, and H-$s$ orbitals were constructed using the \textsc{wannier90} package \cite{souza2001maximally}. For the magnetic-moment calculations, we used a $20\times20\times20$ $k$-mesh.
To demonstrate the interfacial effect in MnTe films that are usually grown on the InP substrate,  we constructed a slab model consisting of 14 MnTe layers (7 unit cells) and 8 InP layers. Atomic positions were fully relaxed at fixed until the force norm on each ion was below 0.01~eV/\AA.

To isolate surface layers from bulk, we projected the Berry curvature onto each atomic layer ($l$) \cite{Varnava2018,Zhao2024}:
\begin{equation}
  \sigma_{xy}(l) = \frac{e^2}{\hbar} \sum_n \int \frac{d^2 k}{(2\pi)^2}  f(\varepsilon_n) \Omega_{n}^{xy} (\bm{k},l),
\end{equation}
where $\Omega_{n}^{xy} (\bm{k},l)$ is the layer-projected Berry curvature of band $n$,
\begin{equation}
  \Omega_{n}^{xy} (\bm{k},l) = -2 \,\mathrm{Im} \sum_{m \neq n} \frac{\braket{n \vert \hat{P}_l \partial_{k_x} H \vert m} \braket{m \vert \partial_{k_y} H \vert n} }{(\varepsilon_n-\varepsilon_m)^2}.
\end{equation}
Here $\hat{P}_l = \ket{l}\bra{l}$ is the projector onto the $l$-th atomic layer. To tame numerical fluctuations in the bulk region caused by the reduced symmetry of Wannier functions, we applied a sliding-average smoothing procedure along the layer index. Specifically, for the internal bulk layers, each smoothed value is computed as the average over four consecutive layers (one unit cell): $\sigma_{xy}(l)=\frac{1}{4}\sum_{k=0}^{3}\sigma_{xy}(l+k)$. Layers outside this range, including the top surface and interface regions, are not smoothed. 

\begin{figure*}[t]
  \centering
  \includegraphics[width=\linewidth]{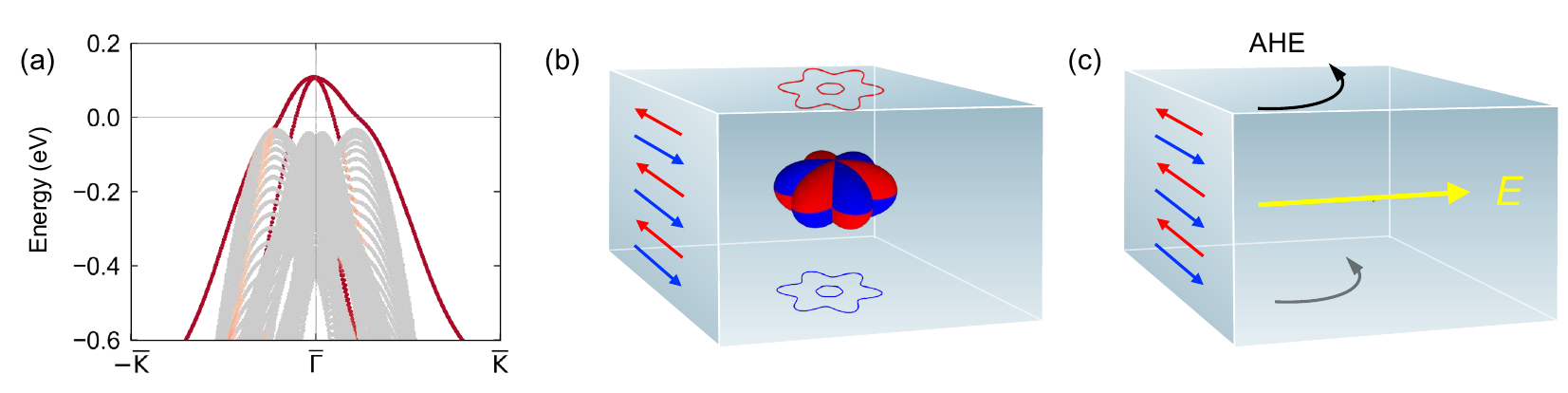}
  \caption{
  (a) Surface band structure with SOC. Red color represents the top surface projection (Te-termination) and gray color indicates the bulk bands in a slab model.  
  (b) Schematics of the spin texture in bulk and surfaces. While bulk bands have a $g$-wave feature, top and bottom surface states are fully spin polarized along $+y$ and $-y$, respectively. The red and blue arrows depict the altermagnetic order.
  (c) Top and bottom surface states induce the same AHE due to the Berry curvature, though they exhibit opposite spin polarization ($\pm S_y$). 
  }
  \label{fig1}
\end{figure*}

\section{Surface States} \label{spin-FS}
Because altermagnetism is fundamentally rooted in symmetry, we first use symmetry analysis to understand the distinction between bulk and surface states. In bulk, magnetic moments oriented along $y$ alternate along the $z$ direction. Two spin sublattices are related by a nonsymmorphic sixfold screw-axis rotation or a mirror reflection, forcing the net spin moment to vanish (in the absence of SOC) with a characteristic $g$-wave-type spin distribution on the Fermi surface [see Fig.~\ref{fig1}(a,b)] \cite{litvin1977spin, chen2024enumeration,jiang2024enumeration,xiao2024spin}. However, these symmetries do not hold on the $xy$ surface and thus, surface states do not necessarily follow the $g$-wave feature, and net surface spin polarization can emerge. 

We indeed observe fully spin-polarized surface bands that are contributed mainly by surface Te-$p_{x,y}$ orbitals. Figure~\ref{fig1}(a) shows the surface states calculated on a slab model with both top and bottom surfaces terminated by Te. Because the single surface Mn layer is spin polarized along $y$, surface states are also spin polarized along the same direction as the surface Mn. Here, we construct the slab model from a tight-binding Hamiltonian using Wannier functions, and fully DFT calculations that consider the surface charge and atomic reconstruction give qualitatively similar results (see details in Section~\ref{interface}). 

The surface has a symmetry $M_x\mathcal{T}$ which transforms spin: 
\begin{equation}
\begin{aligned}
S_x(k_x,k_y) &\rightarrow -S_x(k_x,-k_y), \\
S_y(k_x,k_y) &\rightarrow S_y(k_x,-k_y), \\
S_z(k_x,k_y) &\rightarrow S_z(k_x,-k_y). 
\end{aligned}
\end{equation}
Therefore, $S_x$ vanishes while $S_y$ and $S_z$ are allowed. Since surface Mn atoms exhibit dominant $S_y$ polarization, surface states exhibit the same $S_y$ but a small but nonnegligible $S_z$ as we will discuss in Sections~\ref{AHC_surfaceM} and~\ref{model}. 

In addition, the surface band dispersion is asymmetric regarding $\pm \kk$ when including SOC (see Fig.~\ref{fig1}a) because both inversion and time-reversal symmetries are broken here. Such asymmetry can lead to nonlinear surface transport as discussed in van der Waals antiferromagnets recently~\cite{das2025surface}. 
In ARPES, however, the spectrum appears symmetric, which may arise from spatial averaging over in-plane magnetic domains. We note that recent ARPES measurement is consistent with calculated surface bands~\cite{zhou2026}.

Given that $\alpha$-MnTe is a hole-doped semiconductor, our analysis focuses on the valence bands region. For consistency with experimental observations, we align the Fermi level using the ARPES spectrum along $\overline{\Gamma}$--$\overline{\mathrm{K}}$ as a reference \cite{krempasky2024altermagnetic,hajlaoui2024temperature}. Because the bulk valence top lies at $E^*=-0.03$~eV, we expect surface-dominant transport for chemical potential above $E^*$. 
Notably, the calculated carrier density ($2\times10^{14}$~cm$^{-2}$)
at $E^*$ is much larger than the experimental Hall density of thin films, e.g., $3\times10^{13}$~cm$^{-2}$ in Ref. \cite{gonzalez2023spontaneous} or $5\times10^{13}$~cm$^{-2}$ in Ref. \cite{zhou2026}, indicating that the Fermi level in the experiment is higher than bulk bands. Therefore, it is not surprising that film resistivity is insensitive to thickness in this case. 
Furthermore, the three-dimensional carrier density was reported to be around $10^{19}$ cm$^{-3}$ at 300 K in some bulk samples~\cite{liu2025strain}, which corresponds to a Fermi level 0.02 eV below the bulk edge $E^*$. We note that the two-dimensional carrier density at $E^*$ is equivalent to 200 nm thick bulk, which is larger than the typical thin film thickness (1 $\sim$ 150 nm) reported in literature. 

\section{Surface Anomalous Hall Effect} \label{AHC_thick}

\begin{figure*}[t]
  \centering
  \includegraphics[width=\linewidth]{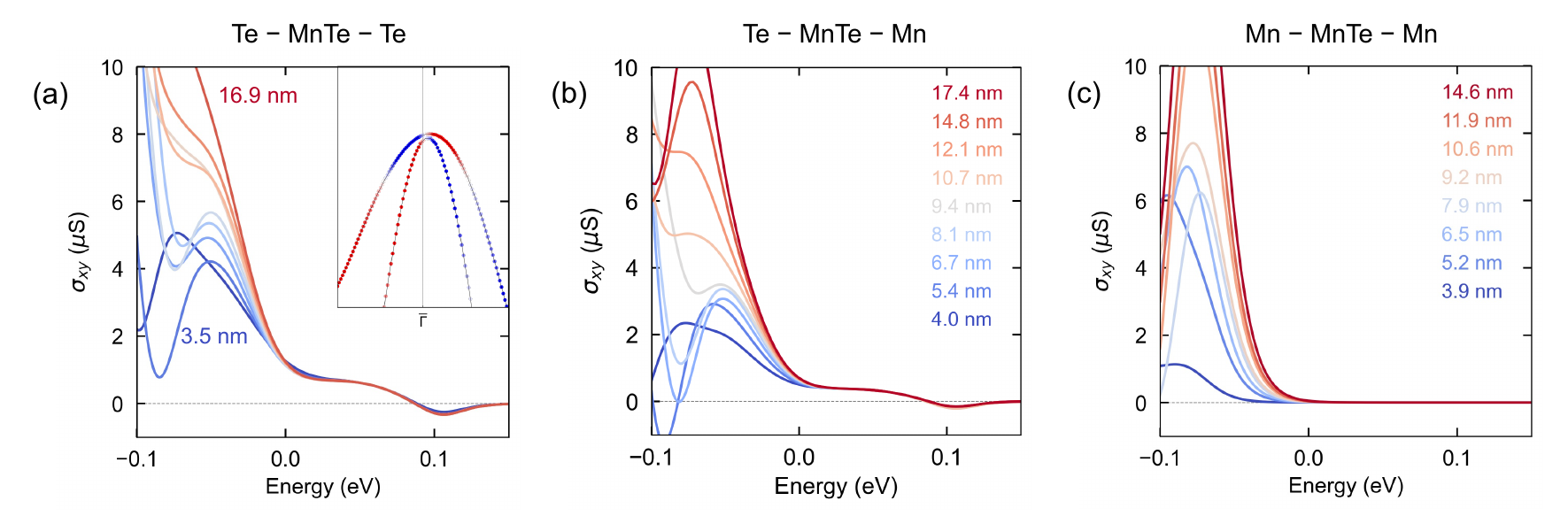}
  \caption{Thickness and energy dependence of the intrinsic anomalous Hall conductance in MnTe slabs with different terminations. (a--c) $\sigma_{xy}^{2\mathrm{D}}(E)$ for slabs with (a) Te/Te, (b) Te/Mn, and (c) Mn/Mn terminations (top/bottom), for several thicknesses. The inset of (a) plots the surface band structure with projected Berry curvature $\Omega_{xy}$ near $\overline{\Gamma}$.}
  \label{fig2}
\end{figure*}

After establishing the surface band structure and spin polarization, we will demonstrate that they generate pronounced AHE in this section. 
We calculate the AHC by integrating the Berry curvature for slabs with different surface terminations and thickness. Our analysis spans the energy window $[-0.1,0.15]$~eV, covering regimes relevant to diverse experimental samples. Figure ~\ref{fig2}(a-c) presents the results for three combinations of surface terminations with film thicknesses ranging from 4 to 17~nm. We considered three combinations of surface terminations: top surface terminated by Mn and bottom surface by Te (Te--MnTe--Mn), both top and bottom surfaces by Te (Te--MnTe--Te), and both top and bottom surfaces by Mn (Mn--MnTe--Mn).

We first take the Te--MnTe--Te slab as an example. Though top and bottom surface bands exhibit opposite polarization in $S_y$, they exhibit the same AHC as summarized by Figs.~\ref{fig1}(b-c). Surface states have two branches distinguished by opposite orbital ($L_z$) textures (see Fig.~\ref{fig:model}). Two surface branches cross at $\Gamma$ in the absence of SOC and open an anti-crossing gap due to SOC. Therefore, two bands exhibit opposite Berry curvature, as shown by the inset of Fig.~\ref{fig2}(a). Consequently, the AHC ($\sigma_{xy}$) is negative when Fermi energy crosses only the upper branch (near 0.1 eV and above) and turns to a nearly positive constant ($\sim 1~\mu$S) after a sign change when shifting to the Fermi level (zero). The nearly constant region indicates that AHC is mainly contributed by the anti-crossing gap. It is natural that AHC is thickness-independent when the Fermi energy is above the bulk valence top.  In addition, given that bulk samples exhibit AHC in the order of magnitude $10^{-2}$ S/cm~\cite{liu2025strain}, the surface AHC ($\sim 1~\mu$S) discussed here is equivalent to a 1000 nm sample if only bulk AHE exists, which is far thicker than those films reported in literature. 

In the asymmetric Te--MnTe--Mn slab [Fig. \ref{fig2}(b)], we find a similar trend of the thickness-independent AHC above $E_F$. Notably, since the Mn-terminated surface state is submerged within the bulk state rather than crossing $E_F$ [(See supplementary materials (SM) Fig. S1 \cite{supplement}], the magnitude of AHC ($\sim 0.5~\mu$S) is half of that in the Te--MnTe--Te slab. This quantitative halving demonstrates that the Te-terminated surfaces act as decoupled parallel transport channels, each contributing equally to the macroscopic AHC, as depicted in Fig. \ref{fig1}(c). Consistent with this picture, the Mn--MnTe--Mn slab yields a vanishing AHC above bulk bands at $E_F$ [Fig. \ref{fig2}(c)].

\begin{figure}[t]
  \centering
  \includegraphics[width=\linewidth]{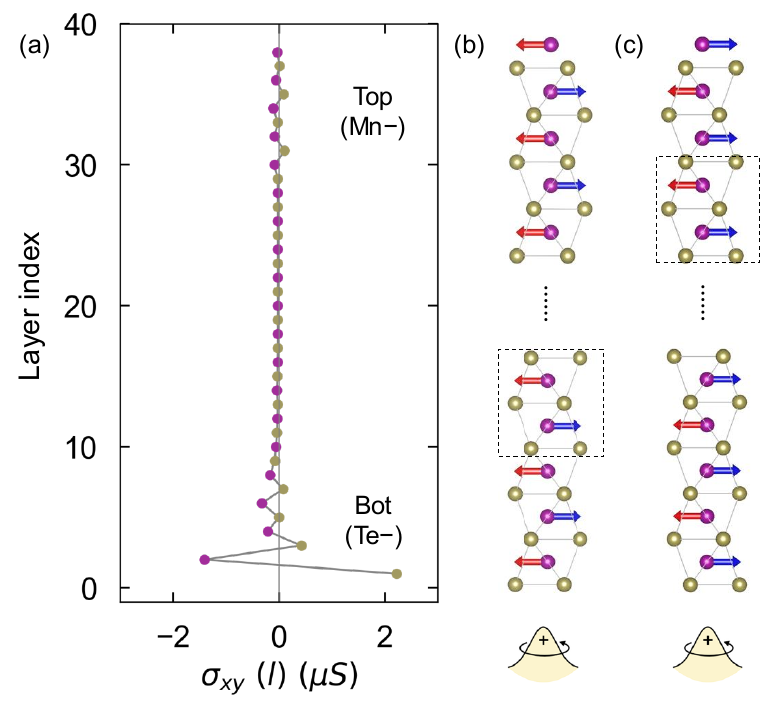}
  \caption{(a) Layer-resolved AHC at $E_F$ in a Te--MnTe--Mn slab (Mn$_{19}$Te$_{19}$). Purple and khaki points represent Mn and Te atomic layers, respectively. (b,c) Corresponding slab structures, Néel-order configuration and corresponding Berry curvature represented by the self-rotating wavepackets. In the dashed boxes, the Mn atoms with the same spin direction share the same local environment. Hence, (b) and (c) share the same bulk Néel order. Because they are symmetric to each other by a $C_{2z}$ rotation, two slabs exhibit the same AHE ($\sigma_{xy}$). }
  \label{fig3}
\end{figure}

Next, we investigate the layer distribution of AHE by constructing a sufficiently thick Te--MnTe--Mn slab to decouple top and bottom surfaces and then projecting AHC onto each atomic layer at $E_F=0$ eV, as shown in Fig.~\ref{fig3}(a). The total AHC originates predominantly from the bottom two unit cells (8 atomic layers) of the Te-terminated surface, while contributions from bulk and the Mn-terminated surface are negligible, consistent with our previous analysis. 

\section{Surface AHE and bulk altermagnetic order} \label{AHC_surfaceM}

Despite the opposite $S_y$ polarization, top and bottom surface bands exhibit the \textit{same} AHE as discussed in Fig.~\ref{fig2}a. Furthermore, we observed the same AHE sign between even and odd numbers of layers, different from the even-odd oscillation of AHE in topological antiferromagnets such as MnBi$_2$Te$_4$~\cite{li2019intrinsic}. 
It is intriguing to ask what determines the AHE polarity for a fixed surface condition in an altermagnet. The answer is the bulk altermagetic order. We will reveal a weak but crucial out-of-plane magnetization ($m_z$) in both bulk and surface, and AHE ($\sigma_{xy}$) manifests this net magnetization. 

We can gain helpful insights by analyzing the bulk magnetic moments. Despite MnTe being nominally considered to have zero magnetization, it actually has a weak but nonvanishing out-of-plane magnetization, which is essential to generate AHE for both bulk and surface. Such a net magnetization is critical for magnetic switching of the Neel vector, similar to the case of Mn$_3$Sn-type chiral antiferromagnets. 
 
\begin{figure}[t]
  \centering
  \includegraphics[width=\linewidth]{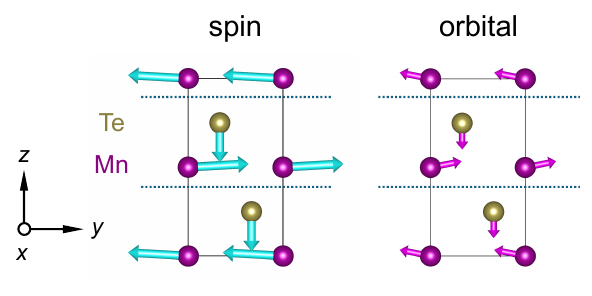}
  \caption{Illustration of the finite out-of-plane ($z$) spin moment and orbital moment in the bulk calculated by DFT. Both Mn sublattices are canting along the same $+z$ direction while all Te atoms exhibit net moments along $-z$. Dashed lines depict two cleaving planes to form Te-terminated surfaces with opposite surface spin sublattices.
  }
  \label{fig4}
\end{figure}

In bulk $\alpha$-MnTe with major spin along the $\pm{y}$ axis, the magnetic moment exhibits an extra component along $z$, which has both spin and orbital contributions, as schematically illustrated in Fig.~\ref{fig4}. For the Mn atom, related spin moment $S_z = 1.1\times10^{-4}$~$\mu_B$ with a canting angle of 0.001371° and orbital moment $L_z=1.2\times10^{-6}$~$\mu_B$ with an canting angle 0.004143°. Further, we identify a finite moment at Te atoms, which aligns purely along $-z$ with $S_z = -1.8\times10^{-4}$~$\mu_B$ and an orbital moment of $L_{z} = -1.8\times10^{-5}$~$\mu_B$, due to hybridization with Mn. 
It is rigorous for the Te moment to align to the $z$ direction because a $z\rightarrow -z$ mirror plane crosses the Te atom in the crystal. 
The total magnetization ($7\times 10^{-5} ~\mu_B$ per formula unit) agrees with the net magnetization ($\sim 5\times 10^{-5}~\mu_B$) reported in experiments \cite{kluczyk2024coexistence,liu2025strain}. 

We highlight that the net $m_z$ is identical between the two spin sublattices, because the two sublattices are related by  $\{C_{2z}|\tau_{c/2}\}$ symmetry that preserves $m_z$. Because of the intimate relation between AHC ($\sigma_{xy}$) and $m_z$, two sublattices also exhibit the same $\sigma_{xy}$. 
This behavior differs from conventional $\mathcal{PT}$ or $\tau \mathcal{T}$-symmetric antiferromagnets, where spin sublattices related by $\mathcal{PT}$ or $\tau \mathcal{T}$ symmetry carry opposite Berry curvature. 
When forming a Te-terminated surface, e.g., by cleaving the bulk from the dashed lines in Fig.~\ref{fig4}, surfaces Te may inherit the bulk Te $m_z$ and exhibit the same AHE polarity regardless of the cleaving plane, i.e., different Mn polarization. In addition, our DFT calculations verify that surface Mn and Te exhibit nonzero $m_z$, which is actually larger than the magnetization inside the bulk. 

We can further rationalize the surface AHE polarity from the surface symmetry. 
We conceive two slabs with opposite surface spin but the same bulk altermagnetic order [Fig. \ref{fig3}(b,c)]. These two slabs can be transformed into each other by a two-fold rotation ($C_{2z}$), which forces two systems to exhibit the same $\sigma_{xy}$. Therefore, the bulk Néel order in MnTe determines surface AHE polarity for a fixed surface termination. 

\section{Effective Model for Surface States} \label{model}

To gain a deeper understanding of surface states and surface AHE, we construct an effective $k\cdot p$ model of MnTe surface states by fitting the energy dispersion and spin and orbital textures of DFT calculations. The symmetry construction of the effective model consists of two steps. In the first step, we neglect the effect of SOC and consider the constraints from the spin space group (SSG)  \cite{litvin1977spin, chen2024enumeration,jiang2024enumeration,xiao2024spin}, which allows for the spin and spatial symmetry operations to be considered independently.  In the second step, we introduce SOC into the effective model.

Without SOC, the bulk MnTe is described by the SSG $P6_3 /^{-1}m^{-1}m^{-1}c^{\infty m}1$, generated by the symmetry operators, $[E \rVert C_{3z}]$, $[E \rVert M_x]$, $[E \rVert \mathcal{I}]$, $[\mathcal{T} \rVert C_{2z}\tau_{c/2}]$, $[M_y \rVert E]$, $[C_{\infty y} \rVert E]$, $[C_{2z}\mathcal{T} \rVert E]$ and translational symmetry \cite{supplement}. Here we use the notation $[\hat{A} \rVert \hat{B}]$ to label the symmetry operator in SSG with $\hat{A}$ denoting a spin symmetry operation and $\hat{B}$ denoting a spatial symmetry operation (here ``spatial" refers to both the orbital and momentum $\kk$). Time reversal $\mathcal{T}$ is an anti-unitary operator and includes a complex conjugate that flips both spin and momentum. $\hat{A}$ can be different from $\hat{B}$ for SSG operator. We note that the spin-only operations ($\hat{B}=E$), exist due to the collinear magnetic order along the $y$ direction. 
On the top surface, the symmetry group is reduced since any symmetry operators related to $z$-directional operation no longer exist, and we find the MnTe surface should be described by spin point group $p^13^1m^{\infty m}1$, with the generators $[E \rVert C_{3z}],[E \rVert M_x],[M_y \rVert E],[C_{\infty y} \rVert E],[C_{2z}\mathcal{T} \rVert E]$, in addition to in-plane translation symmetry. Using these symmetry operators, the effective Hamiltonian can be constructed as
\begin{equation}
\begin{aligned}
&H_{\mathrm{AF}}(\bm{k}) = (a_{0} + a_{2} k^2 + a_{4} k^4)\tau_0\sigma_0 +a_{3}(k_+^3 + k_-^3)\tau_z\sigma_0 \\& + b_{2} [(k_+^2 + k_-^2)\tau_x +i(k_-^2 - k_+^2)\,\tau_y]\sigma_0\\&+b_{4}[(k_+^4 + k_-^4)\tau_x + i(k_+^4 - k_-^4)\,\tau_y]\sigma_0+m\tau_0\sigma_y,
\end{aligned}
\label{eq:H0fullSSG}
\end{equation}
where $k_\pm=k_x +i k_y$, and $\tau$ and $\sigma$ are for the orbital and spin degrees of freedom, respectively. The orbital basis for $\tau$ is chosen to be $p_{\pm} = p_x \pm i p_y$. $a_{0, 2, 3, 4}$ and $b_{2, 4}$ are material parameters independent of spin and $m$ describes the exchange coupling strength between two spin sublattices. We have dropped spin-dependent kinetic terms (See SM \cite{supplement} for more details).   

\begin{figure}[t]
  \centering
  \includegraphics[width=\linewidth]{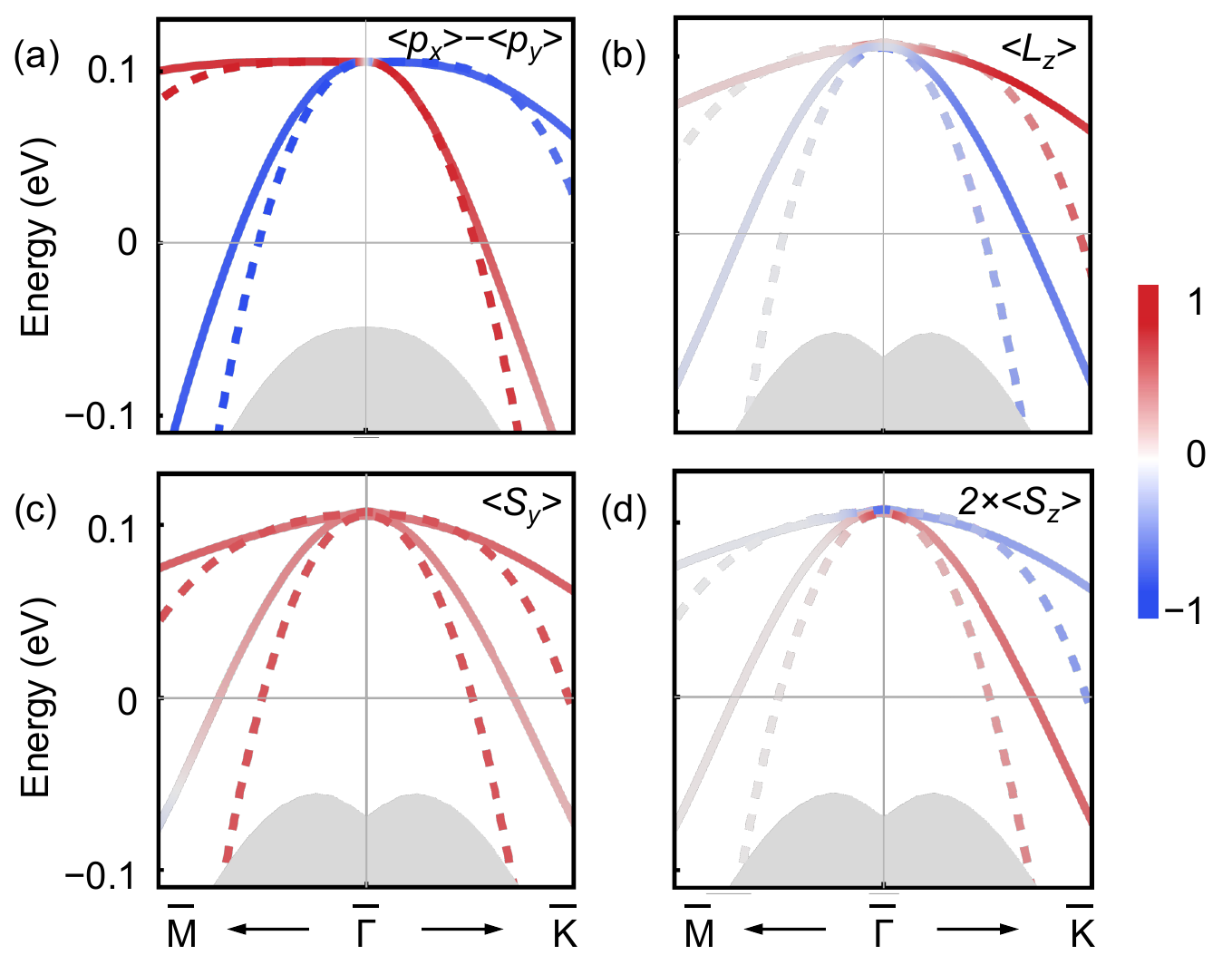}
  \caption{Comparison between band structures along $0.15\overline{\mathrm{M}}$--$\overline{\Gamma}$--0.15$\overline{\mathrm{K}}$ path for DFT calculated (solid) and effective model (dashed) bands, with the projection of (a) orbital hybridization $ \langle p_x \rangle - \langle p_y \rangle$ without SOC, (b) orbital polarization $\langle L_z \rangle$, (c) spin polarization $\langle S_y\rangle$ and (d) twice $\langle S_z\rangle$. (b)-(d) are calculated with SOC. Grey background denotes bulk-like states.
  }
  \label{fig5}
\end{figure}

We consider the eigen-equation
\begin{eqnarray}
    H_{\mathrm{AF}}(\kk)\Psi(\kk) = E\Psi(\kk)
\end{eqnarray}
and note that the $y$ direction spin is a good quantum number as $H_{\mathrm{AF}}(\kk)$ commutes with $\sigma_y$. Consequently, we can choose the eigen wavefunction as 
\begin{eqnarray}
   \Psi_{s}(\kk) = \phi_{s}(\kk)\otimes\chi_s,    
\end{eqnarray}
where $s=\pm$ labels spin up and down, $\chi_s$ is the spin eigenstate of $\sigma_y$, $\sigma_y \chi_s = s \chi_s$, and $\phi_{s}$ is the orbital eigen wavefunction form for spin-$\chi_s$ sector, namely
\begin{eqnarray}
    H_{\mathrm{AF},s}(\kk)\phi_s(\kk) = E\phi_s(\kk)
\end{eqnarray}
with 
\begin{equation}
\begin{aligned}
 &H_{\mathrm{AF},s}(\bm{k}) = (a_0 + sm + a_2 k^2 + a_4 k^4)\,\tau_0\\& +a_{3}(k_+^3 + k_-^3)\tau_z +b_{2} [(k_+^2 + k_-^2)\tau_x 
    + i(k_-^2 - k_+^2)\,\tau_y]\\&+b_{4}[(k_+^4 + k_-^4)\tau_x   + i(k_+^4 - k_-^4)\,\tau_y].
\end{aligned}
\label{eq:H0upSSG}
\end{equation}
The eigen-energies of $H_{\mathrm{AF},s}(\kk)$ are given by
\begin{equation}
\begin{aligned}
&E_{s,\pm}({\bf k}) =a_0 + sm +a_2k^2+a_4k^4 \\
&\pm\sqrt{2}\sqrt{a_3^2k^6+2b_2^2k^4+2b_4^2k^8+(a_3^2+4b_2b_4)k^6\cos(6\theta)}
\label{eq:nosocdisp},
\end{aligned}
\end{equation}
where we label two surface bands as $\pm$. The two bands with higher energies (the spin-$\chi_+$ bands when choosing $m>0$ or the spin-$\chi_-$ bands when choosing $m<0$) can fit well with the surface band dispersion from the DFT calculations around $\overline{\Gamma}$ point ($\kk = 0$) with the parameters listed in SM Table V \cite{supplement}, as shown in Fig. \ref{fig5}(a) and (b) (also see SM Fig. S2 for more details \cite{supplement}). The other two branches of low-energy bands are buried in the bulk energy bands and cannot be resolved.

In Eq.\eqref{eq:nosocdisp}, we note that both the $a_3$ term and the combination of the $b_2$ and $b_4$ terms break the full rotation and give rise to anisotropy in energy dispersion, as shown by the plot of constant energy contour in SM Fig. S2 \cite{supplement}. The $\cos{(6\theta)}$ form in Eq.\eqref{eq:nosocdisp} indicates a six-fold rotational symmetry in the energy dispersion due to the combination of $[E \rVert C_{3z}]$ and $[C_{2z}\mathcal{T} \rVert E]$ in the SSG of MnTe surface. We note the $a_3$ term involves $\tau_z$ while the $b_2$ and $b_4$ terms couple to $\tau_x$ and $\tau_y$. Consequently, we would expect the $a_3$ term to introduce the $p_{\pm}$ orbital polarization in the ${\bf k}$ space, which can be characterized by $z$-component of the orbital angular momentum 
\begin{eqnarray}
L^{(\lambda)}_z =  \langle\tau_{z}\rangle_{\lambda}=\sum_{s=\pm}|\phi^{(\lambda)}_{p_+,s}|^2-|\phi^{(\lambda)}_{p_-,s}|^2    
\end{eqnarray} 
with $\phi^{(\lambda)}_s = (\phi^{(\lambda)}_{p_+,s}, \phi^{(\lambda)}_{p_-,s})^T$ and $\lambda=\pm$ labelling two surface bands, while the $b_2$ and $b_4$ terms introduce non-zero $p_{x,y}$ orbital polarization, defined by 
\begin{eqnarray}
    && \langle p_x\rangle_{\lambda}=\frac{1}{2}\sum_{s=\pm}|\phi^{(\lambda)}_{p_+,s}+\phi^{(\lambda)}_{p_-,s}|^2, \\
    && \langle p_y\rangle_{\lambda}=\frac{1}{2}\sum_{s=\pm}|\phi^{(\lambda)}_{p_+,s}-\phi^{(\lambda)}_{p_-,s}|^2=1-\langle p_x\rangle_\lambda.
\end{eqnarray}
A comparison of energy dispersion and $p$-orbital projection between the effective model and the DFT calculations is shown in Fig. \ref{fig5}(a) for $p_{x,y}$ orbital projection.

\begin{figure*}[t]
  \centering
  \includegraphics[width=\linewidth]{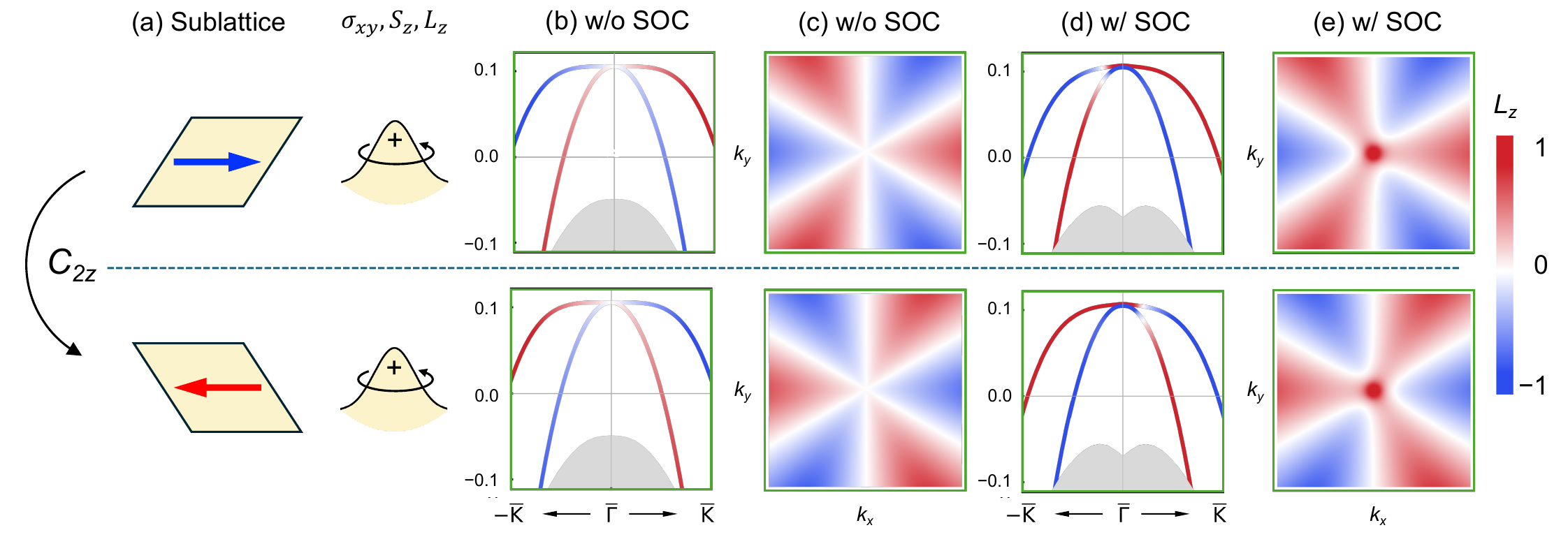}
  \caption{ Band structures and skyrmion-type orbital texture of surface states by the effective model. (a) The magnetic surface sublattice, its out-of-plane moment (spin $S_z$ and orbital $L_z$) and induced AHC $\sigma_{xy}$. (b) Band structure with the $L_z$ projection in the absence of SOC. (c) Distribution of $L_z$ in $\kk$-space without SOC where net $L_z$ vanishes. (d) Band structure with $L_z$ projection after introducing SOC. (e) Distribution of $L_z$ with SOC where net $L_z$ is positive.  Bottom panels are the same as the top ones after transforming the system by $C_{2z}$. The net $L_z$ and $\sigma_{xy}$ appear only after including SOC. Two opposite surface spin sublattices exhibit the same $L_z$ and $\sigma_{xy}$ because of the  $C_{2z}$ symmetry.  
  }\label{fig:model}
\end{figure*}

Next, we introduce SOC into the effective model. As SOC couples spin and orbitals, it reduces the symmetry group in bulk from SSG to magentic space group (MSG) $Cm'c'm$, with generators $C_{2z}\tau_{c/2},M_x\mathcal{T},M_y\tau_{c/2}\mathcal{T}$, in which the symmetry operations act on both the spin and spatial degrees of freedom. At the surface, the absence of any out-of-plane crystal transformation further reduces the MSG to the group with a single $M_x\mathcal{T}$ symmetry operator. In SM Sec.II.C.2, we list all possible symmetry-allowed momentum-independent terms, from which we can identify the relevant SOC terms as 
\begin{eqnarray}
H_{\mathrm{SOC}}=l_z\tau_z\sigma_z+l_y\tau_z\sigma_y,\label{eq:HfullSOC}
\end{eqnarray} 
by comparing the orbital and spin textures as discussed below. We note that the $l_z$ term only shifts the two spin-$s=+$ bands and cannot open a gap, while the $l_y$ term opens a small gap $\sim$ 2 meV, as shown in SM Fig. S3. The $l_y$ term is substantial in inducing Berry curvature around $\overline{\Gamma}$ point. To see that, we also project $H_{\mathrm{SOC}}$ into the spin-$\chi_s$ basis and up to the first order perturbation, we have
\begin{eqnarray}
H^{(1)}_{\mathrm{SOC},s}= s l_y\tau_z, \end{eqnarray} 
and the full Hamiltonian for the spin-$\chi_s$ band is given by 
\begin{equation}
\begin{aligned}\label{eq:H_surface}
    H_{s} &= H_{\mathrm{AF},s} + H^{(1)}_{\mathrm{SOC},s} \\
    &\approx (a_0 + sm + a_2 k^2)\tau_0 + a_{3}(k_+^3 + k_-^3)\tau_z\\ &  +b_{2} [(k_+^2 + k_-^2)\tau_x 
    + i(k_-^2 - k_+^2)\,\tau_y]+sl_y\tau_z,
\end{aligned}
\end{equation}
up to $k^3$ order. When choosing $m>0$, the spin-$\chi_+$ bands have higher energy, so the surface state is determined by the effective model $H_{s=+}$ in Eq. (\ref{eq:H_surface}). In contrast, for $m<0$, the spin-$\chi_-$ bands describe the surface state with the effective model $H_{s=-}$. Thus, the surface bands can be generally described by 
\begin{eqnarray}\label{eq:surface_effective_hamiltonian}
    & H_{\mathrm{surf}} = \text{sign}(m) l_y \tau_z + b_{2} [(k_+^2 + k_-^2)\tau_x + i(k_-^2 - k_+^2)\tau_y]\nonumber \\
    & + a_{3}(k_+^3 + k_-^3)\tau_z + (a_0 + |m| + a_2 k^2) \tau_0.
\end{eqnarray}

The Hamiltonian in Eq. (\ref{eq:surface_effective_hamiltonian}) elucidates the connection between Berry curvature and orbital texture for surface states in MnTe. Consider two surface terminations within a single bulk magnetic domain, characterized by opposite magnetization directions, as shown schematically in Fig. \ref{fig:model}(a). These two surface terminations are related by a two-fold rotation along the $z$ axis ($C_{2z}$), which flips the momentum, $\kk\to-\kk$. As discussed in SM Sec.II.E. \cite{supplement}, the representation matrix for $C_{2z}$ operator is given by $D_{C_{2z}}=-i\tau_0\sigma_z$ on the spin-orbital basis. By acting the $C_{2z}$ operator on the full Hamiltonian Eqs. \eqref{eq:H0fullSSG} and \eqref{eq:HfullSOC} together (See SM Sec.II.E \cite{supplement}), we find that the parameters $m,l_y,a_3$ reverse their signs whereas the parameters $a_0,a_2,a_4,b_4,b_4,l_z$ remain invariant for two surface terminations in Fig. \ref{fig:model}(a). Fig. \ref{fig:model}(c) and \ref{fig:model}(e) show the $L_z$ orbital texture for the case without SOC ($l_y=0$) and with SOC ($l_y\neq 0$), respectively. The resultant dispersion with orbital $p_{\pm}$ polarization from the effective model in Fig. \ref{fig:model}(b,d) is compared with DFT analysis in Fig. \ref{fig5}(b). Without SOC, the orbital polarization exhibits a three-fold angular dependence with alternating positive (red) and negative (blue) lobes in Fig. \ref{fig:model}(c), reflecting the underlying $C_{3z}$ rotation symmetry. This $L_z$ texture originates from the $a_3$ term in Eq. (\ref{eq:surface_effective_hamiltonian}); because $a_3$ flips its sign for two surface terminations, the resulting $L_z$ orbital textures are inverted, as seen by comparing the top and bottom panels in Fig. \ref{fig:model}(b,c). With SOC, the $L_z$ orbital texture remains the same for a large $\kk$, whereas a strong $L_z$-polarization peak appears around $\overline{\Gamma}$ in Fig. \ref{fig:model}(e), as compared to Fig. \ref{fig:model}(c) for the no-SOC case. Two surface bands in Fig. \ref{fig:model}(d) open a small gap at $\overline{\Gamma}$, as shown in SM Fig. S3. Both the small gap and the strong $L_z$ polarization come from the $\text{sign}(m)l_y\tau_z$ term in Eq. (\ref{eq:surface_effective_hamiltonian}). Although both $m$ and $l_y$ exhibit opposite signs for two surface terminations in Fig. \ref{fig:model}(a), their product $\text{sign}(m)l_y$ remains the same, thus giving rise to the same sign of $L_z$ polarization around $\overline{\Gamma}$ in the top and bottom panels of Fig. \ref{fig:model}(e). 

Based on the above understanding of the $L_z$ orbital texture, we next evaluate the Berry curvature distribution, which governs the intrinsic AHE observed in experiments \cite{zhou2026}. In $H_{\mathrm{surf}}$, the $b_2$ term defines a quadratic-band-touching dispersion \cite{sun2009topological}, while the $\tau_z$ term opens a gap at $\overline{\Gamma}$, whose magnitude depends on the parameter $\text{sign}(m) l_z$. If we only keep the terms up to $k^2$ order, the Berry curvature of $H_{\mathrm{surf}}$ in Eq. (\ref{eq:surface_effective_hamiltonian}) can be computed as 
\begin{eqnarray}
\Omega^{\lambda=\pm}_{xy}=\text{sign}(m)\lambda\frac{16l_yb_2^2k^2}{(4b_2^2k^4+l_y^2)^{3/2}}.
\end{eqnarray}
where the $\lambda=\pm$ denotes the two surface bands of $H_{\mathrm{surf}}$. We note that the $p$-orbitals form a skyrmion-like texture around $\overline{\Gamma}$, which is the origin of Berry curvature distribution (See SM Fig. S4 \cite{supplement}).

In this model, $\text{sign}(m) l_z$ determines the AHE polarity. We highlight that $\text{sign}(m)$ refers to surface spin polarization ($\pm S_y$) while $l_y$, which couples the in-plane spin ($\sigma_y$) to the out-of-plane orbital ($\tau_z$), is determined by the local crystal structure. Two surfaces that have opposite surface spin polarization but the same bulk N\'eel order [e.g., Fig.~\ref{fig3}(b,c) or Fig.~\ref{fig:model}] will show opposite signs in both $\text{sign}(m)$ and $l_y$ and thus present the same Berry curvature. This invariance explains why the sign of AHC is robust against which spin sublattice is exposed on the surface. On the other hand, two surfaces that have the same surface spin polarization but opposite bulk N\'eel order will show the same $\text{sign}(m)$ but opposite $l_y$ and thus display opposite AHE. 
Therefore, the surface AHE polarity is determined by the bulk altermagnetic order rather than the surface polarization.

Although the additional SOC term $l_z$ in Eq. (\ref{eq:HfullSOC}) is not substantial for the energy gap between two surface bands and the corresponding AHE (one can confirm the AHC remains non-zero for $l_z=0$), it induces $z$-directional spin texture in the momentum space, as shown in Fig. \ref{fig5}(d) and SM Figs. S4(g,h) \cite{supplement}. Since the $y$-directional exchange coupling $|m|\sim 0.77$ eV is more than three times $l_z\sim 0.2$ eV, the SOC $l_z$ term only induces $z$-directional spin texture perturbatively, while the $y$-directional spin due to magnetization dominates the surface bands, as shown in Fig. \ref{fig5}(c). As $L_z$ polarization remains the same sign for both surface terminations, the spin polarization $S_z$ is also the same, as schematically illustrated in Fig. \ref{fig4}. 

\section{More realistic surfaces and interfaces} \label{interface}

\begin{figure}[tp]
  \centering
  \includegraphics[width=\linewidth]{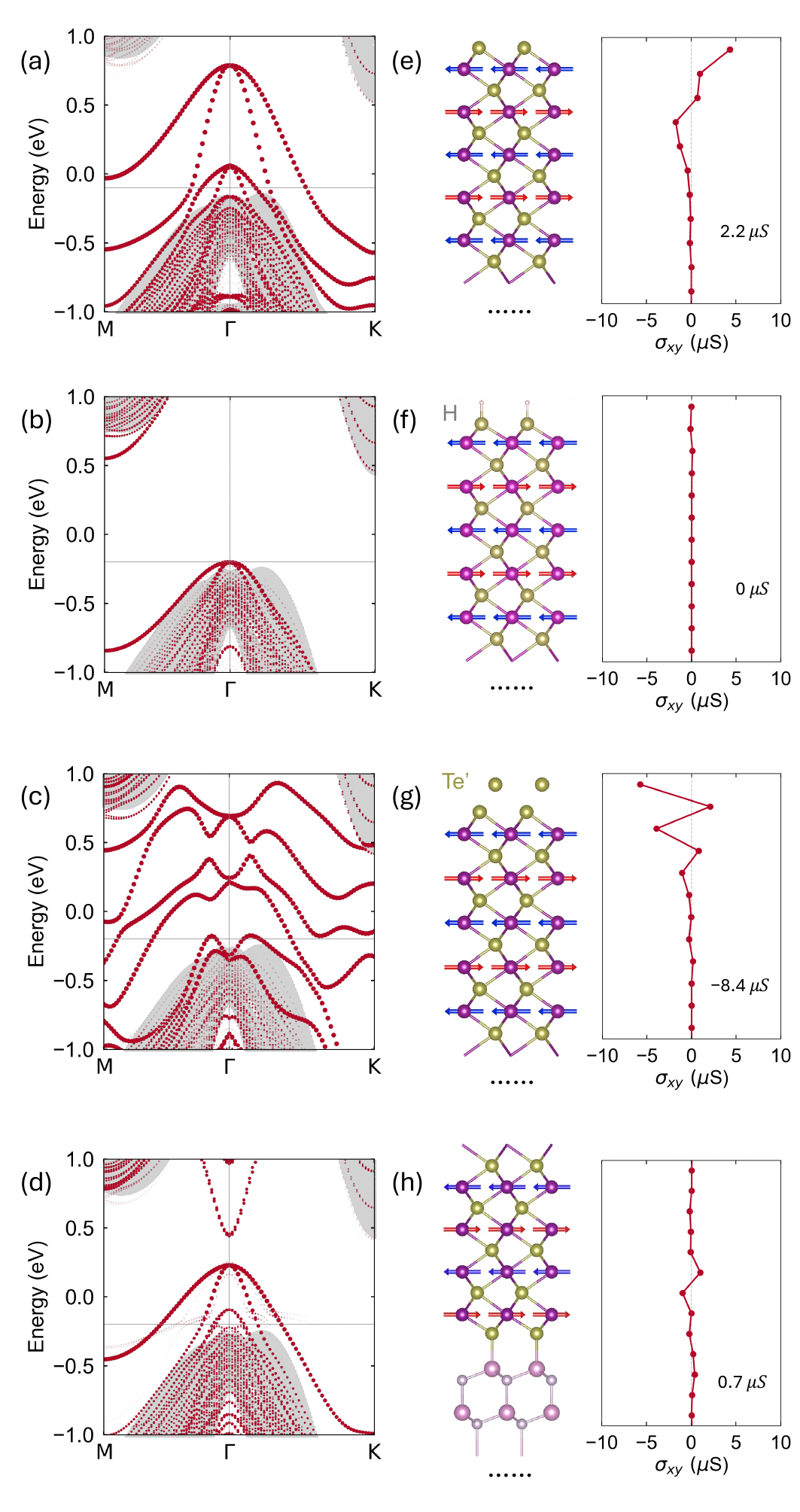}
  \caption{Interface chemistry and its impact on the AHE. (a--d) Band structures for (a) truncated-bulk Te termination, (b) H-passivated Te termination, (c) Te-capped Te termination, and (d) MnTe/InP interface. The projected surface/interface wavefunction weight is highlighted; the grey background indicates bulk-like states. The grey horizontal line marks the Fermi level used for the AHC calculations. (e--h) Corresponding atomic structures (left) and layer-projected AHC contribution $\sigma_{xy}(l)$ (right). The ellipsis denotes interior layers not shown. Red/blue arrows indicate the Néel-order configuration, and the annotated value indicates the summed surface/interface contribution $\sigma_{xy}^{S}$.
  }
  \label{fig7}
\end{figure}

In the previous sections, our calculations relied on a Wannier-based tight-binding Hamiltonian, which assumes an idealized slab geometry. To capture the complexity of realistic experimental samples, we now incorporate the effects of the local chemical environment, including atomic relaxation, charge redistribution, and surface passivation, by constructing four distinct interface configurations.

In bulk MnTe, each Mn (Te) atom forms chemical bonds with six neighboring Te (Mn) atoms, resulting in Mn$^{2+}$ and Te$^{2-}$ charge states. The cleavage of the $xy$ surface creates dangling bonds that host metallic surface states with unpaired electrons, as illustrated in Fig.~\ref{fig7}(a). Such polar surfaces are usually unstable and typically stabilize through mechanisms like surface reconstruction or passivation~\cite{tasker1979stability,noguera2000polar}. Consistent with recent ARPES observations of unoccupied surface states \cite{zhou2026}, we focus on the Te-terminated top surface. In experimental transport devices, this surface is either exposed to air (likely oxidized) or protected by a capping layer.

To model the uncapped/oxidized scenario, we introduce H atoms to saturate dangling bonds. This passivation pushes the surface state into the bulk bands [Fig.~\ref{fig7}(b)], mimicking the suppression of metallic surface states by oxidation. By contrast, protective capping layers often employ chemically compatible tellurides \cite{zhou2026}. These capping layers typically consist of polycrystalline Te chiral chains held together by weak van der Waals forces, minimizing structural disruption to the underlying MnTe. In our simulations, we model this interface by absorbing an additional Te atom (denoted Te$^{\prime}$) onto the Te-terminated surface. We compare two high-symmetry, (0,0) and (2/3,1/3), and adopt the energetically favored configuration ($\Delta E\sim0.085$~eV) for subsequent calculations. In Fig. \ref{fig7}(c), the tellurium capping layer significantly modifies the surface band structure, which can result in a drastically different Berry curvature.

For the bottom surface, we consider an InP(111) substrate. Experimentally, the InP surface is treated with a Te flux prior to MnTe growth, ensuring that Te atoms bond with the topmost In layer. This termination suppresses P-derived surface states and stabilizes the MnTe/InP heterostructure. We analyze the electronic structure of this interface using a local electron-counting rule: each Mn atom donates two electrons to the adjacent Te layers (sharing $\frac{1}{3}$ electron per bond with 6 neighbors), while each In atom in bulk InP donates three electrons to four nearest-neighbor P atoms (sharing $\frac{3}{4}$ electron per bond with 4 neighbors). At the Te--In interface, the charge mismatch results in an excess of approximately $1/4$ electron per interfacial unit. This charge imbalance manifests as a partially filled, hole-like interfacial band, as shown in Fig.~\ref{fig7}(d).

Next, we calculate the layer-resolved AHC for the four interface models, as presented in Fig. \ref{fig7}(e-h), where the Fermi level (grey line) is aligned following the analysis in Section \ref{spin-FS}. Consistent with the idealized slab results in Fig.~\ref{fig3}, AHC is strongly localized within the first few atomic layers near the surface or interface, while the deep interior region contributes only negligible signals. This confirms that $\sigma_{xy}^{S}$ is a well-defined interfacial property, tunable through modification of the local chemical environment.

For the truncated-bulk Te-termination [Fig. \ref{fig7}(e)], the metallic dangling-bond states produce  $\sigma_{xy}^{S}=2.2~\mu$S, with the dominant contribution from the outermost MnTe bilayer. In contrast, H passivation eliminates the dangling-bond state by pushing them into the bulk bands [Fig. \ref{fig7}(b)], thereby suppressing the surface Berry curvature and resulting in a near-zero $\sigma_{xy}^{S}$ [Fig. \ref{fig7}(f)]. This comparison highlights that a metallic surface/interface band near the Fermi level is essential for generating an appreciable surface AHE in MnTe.

The Te-capped termination exhibits the most pronounced modification [Fig. \ref{fig7}(g)]. Hybridization between the top Te layer and the MnTe surface reconstructs the electronic structure near $E_F$, redistributing the Berry curvature over the outermost layers. This leads to a large and sign-reversed surface contribution $\sigma_{xy}^{S}=-8.4~\mu$S. This sensitivity to capping chemistry provides a microscopic explanation for the experimental observation that nominally similar films can exhibit opposite Hall hysteresis loops when protected by a telluride layer. At the MnTe/InP interface [Fig. \ref{fig7}(h)], the hole-like interfacial band discussed above gives rise to a positive $\sigma_{xy}^{S}$, again predominantly localized within the first few layers of MnTe.

These results indicate that the measured Hall signal in MnTe heterostructures can be understood as a superposition of two surface/interface contributions, whose signs and magnitudes are set by termination and local chemistry. Specifically, in an uncapped thin film grown on an InP substrate, the net AHC is dominated by the positive contribution from the bottom interface. When the top surface is Te-capped, the two interfaces contribute with opposite signs, leading to complicated hysteresis loops. Depending on their relative magnitudes, the net AHC can be significantly reduced or even reversed, reconciling the diverse transport behaviors reported in the literature.

In addition, we focus on the intrinsic AHE from the clean surface band structure in this present work. Extrinsic effects like skew scattering or side jump due to disorders \cite{RevModPhys.82.1539} may coexist with intrinsic AHE in experiments. Nevertheless, extrinsic AHE will also respect the same surface symmetry as the intrinsic effect and thus follow the same influence from the bulk magnetic order. 

\section{Summary} \label{discussion}
In summary, we demonstrate that emergent surface states generate a large AHE driven by their strong Berry curvature near the Fermi level, rationalizing the unusual AHE and resistivity observed in MnTe thin films. 
Though it is sensitive to the local surface chemistry condition, the surface AHE has a deep bulk origin. 
Building upon the deterministic relationship between the surface AHC and the bulk altermagnetic order, we propose utilizing the sign of the surface AHE as an unambiguous electrical probe for the bulk Néel order. 
These findings not only provide a microscopic resolution to the transport signals reported experimentally but also establish surface and interface engineering as a critical avenue for tailoring altermagnetic transport phenomena in future spintronic devices.

\section{Acknowledgments}
We thank the fruitful discussion with Cuizu Chang, Lingjie Zhou, Chunhui Du, Zijie Yan, Hongtao Rong, Pu Xiao, Kamal Das, and Diana Golovanova. B.Y. and C.X.L. acknowledge the National Science Foundation through the Penn State Materials Research Science and Engineering Center (MRSEC) DMR 2011839. B.Y. acknowledges the financial support by the Israel Science Foundation (ISF  No. 2974/23).

\FloatBarrier


\begin{thebibliography}{49}%
\makeatletter
\providecommand \@ifxundefined [1]{%
 \@ifx{#1\undefined}
}%
\providecommand \@ifnum [1]{%
 \ifnum #1\expandafter \@firstoftwo
 \else \expandafter \@secondoftwo
 \fi
}%
\providecommand \@ifx [1]{%
 \ifx #1\expandafter \@firstoftwo
 \else \expandafter \@secondoftwo
 \fi
}%
\providecommand \natexlab [1]{#1}%
\providecommand \enquote  [1]{``#1''}%
\providecommand \bibnamefont  [1]{#1}%
\providecommand \bibfnamefont [1]{#1}%
\providecommand \citenamefont [1]{#1}%
\providecommand \href@noop [0]{\@secondoftwo}%
\providecommand \href [0]{\begingroup \@sanitize@url \@href}%
\providecommand \@href[1]{\@@startlink{#1}\@@href}%
\providecommand \@@href[1]{\endgroup#1\@@endlink}%
\providecommand \@sanitize@url [0]{\catcode `\\12\catcode `\$12\catcode `\&12\catcode `\#12\catcode `\^12\catcode `\_12\catcode `\%12\relax}%
\providecommand \@@startlink[1]{}%
\providecommand \@@endlink[0]{}%
\providecommand \url  [0]{\begingroup\@sanitize@url \@url }%
\providecommand \@url [1]{\endgroup\@href {#1}{\urlprefix }}%
\providecommand \urlprefix  [0]{URL }%
\providecommand \Eprint [0]{\href }%
\providecommand \doibase [0]{https://doi.org/}%
\providecommand \selectlanguage [0]{\@gobble}%
\providecommand \bibinfo  [0]{\@secondoftwo}%
\providecommand \bibfield  [0]{\@secondoftwo}%
\providecommand \translation [1]{[#1]}%
\providecommand \BibitemOpen [0]{}%
\providecommand \bibitemStop [0]{}%
\providecommand \bibitemNoStop [0]{.\EOS\space}%
\providecommand \EOS [0]{\spacefactor3000\relax}%
\providecommand \BibitemShut  [1]{\csname bibitem#1\endcsname}%
\let\auto@bib@innerbib\@empty
\bibitem [{\citenamefont {{\v{S}}mejkal}\ \emph {et~al.}(2020)\citenamefont {{\v{S}}mejkal}, \citenamefont {Gonz{\'a}lez-Hern{\'a}ndez}, \citenamefont {Jungwirth},\ and\ \citenamefont {Sinova}}]{vsmejkal2020crystal}%
  \BibitemOpen
  \bibfield  {author} {\bibinfo {author} {\bibfnamefont {L.}~\bibnamefont {{\v{S}}mejkal}}, \bibinfo {author} {\bibfnamefont {R.}~\bibnamefont {Gonz{\'a}lez-Hern{\'a}ndez}}, \bibinfo {author} {\bibfnamefont {T.}~\bibnamefont {Jungwirth}},\ and\ \bibinfo {author} {\bibfnamefont {J.}~\bibnamefont {Sinova}},\ }\bibfield  {title} {\bibinfo {title} {Crystal time-reversal symmetry breaking and spontaneous hall effect in collinear antiferromagnets},\ }\href {https://www.science.org/doi/10.1126/sciadv.aaz8809} {\bibfield  {journal} {\bibinfo  {journal} {Science advances}\ }\textbf {\bibinfo {volume} {6}},\ \bibinfo {pages} {eaaz8809} (\bibinfo {year} {2020})}\BibitemShut {NoStop}%
\bibitem [{\citenamefont {Yuan}\ \emph {et~al.}(2020)\citenamefont {Yuan}, \citenamefont {Wang}, \citenamefont {Luo}, \citenamefont {Rashba},\ and\ \citenamefont {Zunger}}]{Yuan2020}%
  \BibitemOpen
  \bibfield  {author} {\bibinfo {author} {\bibfnamefont {L.-D.}\ \bibnamefont {Yuan}}, \bibinfo {author} {\bibfnamefont {Z.}~\bibnamefont {Wang}}, \bibinfo {author} {\bibfnamefont {J.-W.}\ \bibnamefont {Luo}}, \bibinfo {author} {\bibfnamefont {E.~I.}\ \bibnamefont {Rashba}},\ and\ \bibinfo {author} {\bibfnamefont {A.}~\bibnamefont {Zunger}},\ }\bibfield  {title} {\bibinfo {title} {Giant momentum-dependent spin splitting in centrosymmetric low-$z$ antiferromagnets},\ }\href {https://doi.org/10.1103/PhysRevB.102.014422} {\bibfield  {journal} {\bibinfo  {journal} {Phys. Rev. B}\ }\textbf {\bibinfo {volume} {102}},\ \bibinfo {pages} {014422} (\bibinfo {year} {2020})}\BibitemShut {NoStop}%
\bibitem [{\citenamefont {Ma}\ \emph {et~al.}(2021)\citenamefont {Ma}, \citenamefont {Hu}, \citenamefont {Li}, \citenamefont {Liu}, \citenamefont {Yao}, \citenamefont {Jia},\ and\ \citenamefont {Liu}}]{ma2021multifunctional}%
  \BibitemOpen
  \bibfield  {author} {\bibinfo {author} {\bibfnamefont {H.-Y.}\ \bibnamefont {Ma}}, \bibinfo {author} {\bibfnamefont {M.}~\bibnamefont {Hu}}, \bibinfo {author} {\bibfnamefont {N.}~\bibnamefont {Li}}, \bibinfo {author} {\bibfnamefont {J.}~\bibnamefont {Liu}}, \bibinfo {author} {\bibfnamefont {W.}~\bibnamefont {Yao}}, \bibinfo {author} {\bibfnamefont {J.-F.}\ \bibnamefont {Jia}},\ and\ \bibinfo {author} {\bibfnamefont {J.}~\bibnamefont {Liu}},\ }\bibfield  {title} {\bibinfo {title} {Multifunctional antiferromagnetic materials with giant piezomagnetism and noncollinear spin current},\ }\href {https://www.nature.com/articles/s41467-021-23127-7} {\bibfield  {journal} {\bibinfo  {journal} {Nature communications}\ }\textbf {\bibinfo {volume} {12}},\ \bibinfo {pages} {2846} (\bibinfo {year} {2021})}\BibitemShut {NoStop}%
\bibitem [{\citenamefont {{\v{S}}mejkal}\ \emph {et~al.}(2022{\natexlab{a}})\citenamefont {{\v{S}}mejkal}, \citenamefont {Sinova},\ and\ \citenamefont {Jungwirth}}]{vsmejkal2022beyond}%
  \BibitemOpen
  \bibfield  {author} {\bibinfo {author} {\bibfnamefont {L.}~\bibnamefont {{\v{S}}mejkal}}, \bibinfo {author} {\bibfnamefont {J.}~\bibnamefont {Sinova}},\ and\ \bibinfo {author} {\bibfnamefont {T.}~\bibnamefont {Jungwirth}},\ }\bibfield  {title} {\bibinfo {title} {Beyond conventional ferromagnetism and antiferromagnetism: A phase with nonrelativistic spin and crystal rotation symmetry},\ }\href {https://journals.aps.org/prx/abstract/10.1103/PhysRevX.12.031042} {\bibfield  {journal} {\bibinfo  {journal} {Physical Review X}\ }\textbf {\bibinfo {volume} {12}},\ \bibinfo {pages} {031042} (\bibinfo {year} {2022}{\natexlab{a}})}\BibitemShut {NoStop}%
\bibitem [{\citenamefont {{\v{S}}mejkal}\ \emph {et~al.}(2022{\natexlab{b}})\citenamefont {{\v{S}}mejkal}, \citenamefont {Sinova},\ and\ \citenamefont {Jungwirth}}]{vsmejkal2022emerging}%
  \BibitemOpen
  \bibfield  {author} {\bibinfo {author} {\bibfnamefont {L.}~\bibnamefont {{\v{S}}mejkal}}, \bibinfo {author} {\bibfnamefont {J.}~\bibnamefont {Sinova}},\ and\ \bibinfo {author} {\bibfnamefont {T.}~\bibnamefont {Jungwirth}},\ }\bibfield  {title} {\bibinfo {title} {Emerging research landscape of altermagnetism},\ }\href {https://journals.aps.org/prx/abstract/10.1103/PhysRevX.12.040501} {\bibfield  {journal} {\bibinfo  {journal} {Physical Review X}\ }\textbf {\bibinfo {volume} {12}},\ \bibinfo {pages} {040501} (\bibinfo {year} {2022}{\natexlab{b}})}\BibitemShut {NoStop}%
\bibitem [{\citenamefont {Song}\ \emph {et~al.}(2025)\citenamefont {Song}, \citenamefont {Bai}, \citenamefont {Zhou}, \citenamefont {Han}, \citenamefont {Reichlova}, \citenamefont {Dil}, \citenamefont {Liu}, \citenamefont {Chen},\ and\ \citenamefont {Pan}}]{song2025altermagnets}%
  \BibitemOpen
  \bibfield  {author} {\bibinfo {author} {\bibfnamefont {C.}~\bibnamefont {Song}}, \bibinfo {author} {\bibfnamefont {H.}~\bibnamefont {Bai}}, \bibinfo {author} {\bibfnamefont {Z.}~\bibnamefont {Zhou}}, \bibinfo {author} {\bibfnamefont {L.}~\bibnamefont {Han}}, \bibinfo {author} {\bibfnamefont {H.}~\bibnamefont {Reichlova}}, \bibinfo {author} {\bibfnamefont {J.~H.}\ \bibnamefont {Dil}}, \bibinfo {author} {\bibfnamefont {J.}~\bibnamefont {Liu}}, \bibinfo {author} {\bibfnamefont {X.}~\bibnamefont {Chen}},\ and\ \bibinfo {author} {\bibfnamefont {F.}~\bibnamefont {Pan}},\ }\bibfield  {title} {\bibinfo {title} {Altermagnets as a new class of functional materials},\ }\href {https://www.nature.com/articles/s41578-025-00779-1} {\bibfield  {journal} {\bibinfo  {journal} {Nature Reviews Materials}\ ,\ \bibinfo {pages} {1}} (\bibinfo {year} {2025})}\BibitemShut {NoStop}%
\bibitem [{\citenamefont {Amin}\ \emph {et~al.}(2024)\citenamefont {Amin}, \citenamefont {Dal~Din}, \citenamefont {Golias}, \citenamefont {Niu}, \citenamefont {Zakharov}, \citenamefont {Fromage}, \citenamefont {Fields}, \citenamefont {Heywood}, \citenamefont {Cousins}, \citenamefont {Maccherozzi} \emph {et~al.}}]{amin2024nanoscale}%
  \BibitemOpen
  \bibfield  {author} {\bibinfo {author} {\bibfnamefont {O.}~\bibnamefont {Amin}}, \bibinfo {author} {\bibfnamefont {A.}~\bibnamefont {Dal~Din}}, \bibinfo {author} {\bibfnamefont {E.}~\bibnamefont {Golias}}, \bibinfo {author} {\bibfnamefont {Y.}~\bibnamefont {Niu}}, \bibinfo {author} {\bibfnamefont {A.}~\bibnamefont {Zakharov}}, \bibinfo {author} {\bibfnamefont {S.}~\bibnamefont {Fromage}}, \bibinfo {author} {\bibfnamefont {C.}~\bibnamefont {Fields}}, \bibinfo {author} {\bibfnamefont {S.}~\bibnamefont {Heywood}}, \bibinfo {author} {\bibfnamefont {R.}~\bibnamefont {Cousins}}, \bibinfo {author} {\bibfnamefont {F.}~\bibnamefont {Maccherozzi}}, \emph {et~al.},\ }\bibfield  {title} {\bibinfo {title} {Nanoscale imaging and control of altermagnetism in mnte},\ }\href {https://www.nature.com/articles/s41586-024-08234-x} {\bibfield  {journal} {\bibinfo  {journal} {Nature}\ }\textbf {\bibinfo {volume} {636}},\ \bibinfo {pages} {348} (\bibinfo {year} {2024})}\BibitemShut {NoStop}%
\bibitem [{\citenamefont {Mazin}(2022)}]{Mazin2022}%
  \BibitemOpen
  \bibfield  {author} {\bibinfo {author} {\bibfnamefont {I.}~\bibnamefont {Mazin}} (\bibinfo {collaboration} {The PRX Editors}),\ }\bibfield  {title} {\bibinfo {title} {Editorial: Altermagnetism---a new punch line of fundamental magnetism},\ }\href {https://doi.org/10.1103/PhysRevX.12.040002} {\bibfield  {journal} {\bibinfo  {journal} {Phys. Rev. X}\ }\textbf {\bibinfo {volume} {12}},\ \bibinfo {pages} {040002} (\bibinfo {year} {2022})}\BibitemShut {NoStop}%
\bibitem [{\citenamefont {Wu}\ \emph {et~al.}(2007)\citenamefont {Wu}, \citenamefont {Sun}, \citenamefont {Fradkin},\ and\ \citenamefont {Zhang}}]{Wu2007}%
  \BibitemOpen
  \bibfield  {author} {\bibinfo {author} {\bibfnamefont {C.}~\bibnamefont {Wu}}, \bibinfo {author} {\bibfnamefont {K.}~\bibnamefont {Sun}}, \bibinfo {author} {\bibfnamefont {E.}~\bibnamefont {Fradkin}},\ and\ \bibinfo {author} {\bibfnamefont {S.-C.}\ \bibnamefont {Zhang}},\ }\bibfield  {title} {\bibinfo {title} {Fermi liquid instabilities in the spin channel},\ }\href {https://doi.org/10.1103/PhysRevB.75.115103} {\bibfield  {journal} {\bibinfo  {journal} {Phys. Rev. B}\ }\textbf {\bibinfo {volume} {75}},\ \bibinfo {pages} {115103} (\bibinfo {year} {2007})}\BibitemShut {NoStop}%
\bibitem [{\citenamefont {Liu}\ \emph {et~al.}(2025{\natexlab{a}})\citenamefont {Liu}, \citenamefont {Dai},\ and\ \citenamefont {Bl{\"u}gel}}]{liu2025different}%
  \BibitemOpen
  \bibfield  {author} {\bibinfo {author} {\bibfnamefont {Q.}~\bibnamefont {Liu}}, \bibinfo {author} {\bibfnamefont {X.}~\bibnamefont {Dai}},\ and\ \bibinfo {author} {\bibfnamefont {S.}~\bibnamefont {Bl{\"u}gel}},\ }\bibfield  {title} {\bibinfo {title} {Different facets of unconventional magnetism},\ }\href {https://www.nature.com/articles/s41567-024-02750-3} {\bibfield  {journal} {\bibinfo  {journal} {Nature Physics}\ }\textbf {\bibinfo {volume} {21}},\ \bibinfo {pages} {329} (\bibinfo {year} {2025}{\natexlab{a}})}\BibitemShut {NoStop}%
\bibitem [{\citenamefont {Nakatsuji}\ \emph {et~al.}(2015)\citenamefont {Nakatsuji}, \citenamefont {Kiyohara},\ and\ \citenamefont {Higo}}]{nakatsuji2015large}%
  \BibitemOpen
  \bibfield  {author} {\bibinfo {author} {\bibfnamefont {S.}~\bibnamefont {Nakatsuji}}, \bibinfo {author} {\bibfnamefont {N.}~\bibnamefont {Kiyohara}},\ and\ \bibinfo {author} {\bibfnamefont {T.}~\bibnamefont {Higo}},\ }\bibfield  {title} {\bibinfo {title} {Large anomalous hall effect in a non-collinear antiferromagnet at room temperature},\ }\href {https://www.nature.com/articles/nature19416} {\bibfield  {journal} {\bibinfo  {journal} {Nature}\ }\textbf {\bibinfo {volume} {527}},\ \bibinfo {pages} {212} (\bibinfo {year} {2015})}\BibitemShut {NoStop}%
\bibitem [{\citenamefont {Nayak}\ \emph {et~al.}(2016)\citenamefont {Nayak}, \citenamefont {Fischer}, \citenamefont {Sun}, \citenamefont {Yan}, \citenamefont {Karel}, \citenamefont {Komarek}, \citenamefont {Shekhar}, \citenamefont {Kumar}, \citenamefont {Schnelle}, \citenamefont {K{\"u}bler} \emph {et~al.}}]{nayak2016large}%
  \BibitemOpen
  \bibfield  {author} {\bibinfo {author} {\bibfnamefont {A.~K.}\ \bibnamefont {Nayak}}, \bibinfo {author} {\bibfnamefont {J.~E.}\ \bibnamefont {Fischer}}, \bibinfo {author} {\bibfnamefont {Y.}~\bibnamefont {Sun}}, \bibinfo {author} {\bibfnamefont {B.}~\bibnamefont {Yan}}, \bibinfo {author} {\bibfnamefont {J.}~\bibnamefont {Karel}}, \bibinfo {author} {\bibfnamefont {A.~C.}\ \bibnamefont {Komarek}}, \bibinfo {author} {\bibfnamefont {C.}~\bibnamefont {Shekhar}}, \bibinfo {author} {\bibfnamefont {N.}~\bibnamefont {Kumar}}, \bibinfo {author} {\bibfnamefont {W.}~\bibnamefont {Schnelle}}, \bibinfo {author} {\bibfnamefont {J.}~\bibnamefont {K{\"u}bler}}, \emph {et~al.},\ }\bibfield  {title} {\bibinfo {title} {Large anomalous hall effect driven by a nonvanishing berry curvature in the noncolinear antiferromagnet mn3ge},\ }\href {https://www.science.org/doi/full/10.1126/sciadv.1501870} {\bibfield  {journal} {\bibinfo  {journal} {Science advances}\ }\textbf {\bibinfo {volume} {2}},\ \bibinfo {pages} {e1501870} (\bibinfo
  {year} {2016})}\BibitemShut {NoStop}%
\bibitem [{\citenamefont {Zelezny}\ \emph {et~al.}(2017)\citenamefont {Zelezny}, \citenamefont {Zhang}, \citenamefont {Felser},\ and\ \citenamefont {Yan}}]{zelezny2017}%
  \BibitemOpen
  \bibfield  {author} {\bibinfo {author} {\bibfnamefont {J.}~\bibnamefont {Zelezny}}, \bibinfo {author} {\bibfnamefont {Y.}~\bibnamefont {Zhang}}, \bibinfo {author} {\bibfnamefont {C.}~\bibnamefont {Felser}},\ and\ \bibinfo {author} {\bibfnamefont {B.}~\bibnamefont {Yan}},\ }\bibfield  {title} {\bibinfo {title} {Spin-polarized current in noncollinear antiferromagnets},\ }\href {https://doi.org/10.1103/PhysRevLett.119.187204} {\bibfield  {journal} {\bibinfo  {journal} {Phys. Rev. Lett.}\ }\textbf {\bibinfo {volume} {119}},\ \bibinfo {pages} {187204} (\bibinfo {year} {2017})}\BibitemShut {NoStop}%
\bibitem [{\citenamefont {Zhang}\ \emph {et~al.}(2017)\citenamefont {Zhang}, \citenamefont {Sun}, \citenamefont {Yang}, \citenamefont {\ifmmode~\check{Z}\else \v{Z}\fi{}elezn\'y}, \citenamefont {Parkin}, \citenamefont {Felser},\ and\ \citenamefont {Yan}}]{Zhang2017strong}%
  \BibitemOpen
  \bibfield  {author} {\bibinfo {author} {\bibfnamefont {Y.}~\bibnamefont {Zhang}}, \bibinfo {author} {\bibfnamefont {Y.}~\bibnamefont {Sun}}, \bibinfo {author} {\bibfnamefont {H.}~\bibnamefont {Yang}}, \bibinfo {author} {\bibfnamefont {J.}~\bibnamefont {\ifmmode~\check{Z}\else \v{Z}\fi{}elezn\'y}}, \bibinfo {author} {\bibfnamefont {S.~P.~P.}\ \bibnamefont {Parkin}}, \bibinfo {author} {\bibfnamefont {C.}~\bibnamefont {Felser}},\ and\ \bibinfo {author} {\bibfnamefont {B.}~\bibnamefont {Yan}},\ }\bibfield  {title} {\bibinfo {title} {Strong anisotropic anomalous hall effect and spin hall effect in the chiral antiferromagnetic compounds ${\mathrm{mn}}_{3}x$ ($x=\mathrm{Ge}$, sn, ga, ir, rh, and pt)},\ }\href {https://doi.org/10.1103/PhysRevB.95.075128} {\bibfield  {journal} {\bibinfo  {journal} {Phys. Rev. B}\ }\textbf {\bibinfo {volume} {95}},\ \bibinfo {pages} {075128} (\bibinfo {year} {2017})}\BibitemShut {NoStop}%
\bibitem [{\citenamefont {Lee}\ \emph {et~al.}(2024)\citenamefont {Lee}, \citenamefont {Lee}, \citenamefont {Jung}, \citenamefont {Jung}, \citenamefont {Kim}, \citenamefont {Lee}, \citenamefont {Seok}, \citenamefont {Kim}, \citenamefont {Park}, \citenamefont {\ifmmode~\check{S}\else \v{S}\fi{}mejkal}, \citenamefont {Kang},\ and\ \citenamefont {Kim}}]{Lee2024prl}%
  \BibitemOpen
  \bibfield  {author} {\bibinfo {author} {\bibfnamefont {S.}~\bibnamefont {Lee}}, \bibinfo {author} {\bibfnamefont {S.}~\bibnamefont {Lee}}, \bibinfo {author} {\bibfnamefont {S.}~\bibnamefont {Jung}}, \bibinfo {author} {\bibfnamefont {J.}~\bibnamefont {Jung}}, \bibinfo {author} {\bibfnamefont {D.}~\bibnamefont {Kim}}, \bibinfo {author} {\bibfnamefont {Y.}~\bibnamefont {Lee}}, \bibinfo {author} {\bibfnamefont {B.}~\bibnamefont {Seok}}, \bibinfo {author} {\bibfnamefont {J.}~\bibnamefont {Kim}}, \bibinfo {author} {\bibfnamefont {B.~G.}\ \bibnamefont {Park}}, \bibinfo {author} {\bibfnamefont {L.}~\bibnamefont {\ifmmode~\check{S}\else \v{S}\fi{}mejkal}}, \bibinfo {author} {\bibfnamefont {C.-J.}\ \bibnamefont {Kang}},\ and\ \bibinfo {author} {\bibfnamefont {C.}~\bibnamefont {Kim}},\ }\bibfield  {title} {\bibinfo {title} {Broken kramers degeneracy in altermagnetic mnte},\ }\href {https://doi.org/10.1103/PhysRevLett.132.036702} {\bibfield  {journal} {\bibinfo  {journal} {Phys. Rev. Lett.}\ }\textbf {\bibinfo {volume}
  {132}},\ \bibinfo {pages} {036702} (\bibinfo {year} {2024})}\BibitemShut {NoStop}%
\bibitem [{\citenamefont {Fedchenko}\ \emph {et~al.}(2024)\citenamefont {Fedchenko}, \citenamefont {Min{\'a}r}, \citenamefont {Akashdeep}, \citenamefont {D’Souza}, \citenamefont {Vasilyev}, \citenamefont {Tkach}, \citenamefont {Odenbreit}, \citenamefont {Nguyen}, \citenamefont {Kutnyakhov}, \citenamefont {Wind} \emph {et~al.}}]{fedchenko2024observation}%
  \BibitemOpen
  \bibfield  {author} {\bibinfo {author} {\bibfnamefont {O.}~\bibnamefont {Fedchenko}}, \bibinfo {author} {\bibfnamefont {J.}~\bibnamefont {Min{\'a}r}}, \bibinfo {author} {\bibfnamefont {A.}~\bibnamefont {Akashdeep}}, \bibinfo {author} {\bibfnamefont {S.~W.}\ \bibnamefont {D’Souza}}, \bibinfo {author} {\bibfnamefont {D.}~\bibnamefont {Vasilyev}}, \bibinfo {author} {\bibfnamefont {O.}~\bibnamefont {Tkach}}, \bibinfo {author} {\bibfnamefont {L.}~\bibnamefont {Odenbreit}}, \bibinfo {author} {\bibfnamefont {Q.}~\bibnamefont {Nguyen}}, \bibinfo {author} {\bibfnamefont {D.}~\bibnamefont {Kutnyakhov}}, \bibinfo {author} {\bibfnamefont {N.}~\bibnamefont {Wind}}, \emph {et~al.},\ }\bibfield  {title} {\bibinfo {title} {Observation of time-reversal symmetry breaking in the band structure of altermagnetic ruo2},\ }\href {https://www.science.org/doi/10.1126/sciadv.adj4883} {\bibfield  {journal} {\bibinfo  {journal} {Science advances}\ }\textbf {\bibinfo {volume} {10}},\ \bibinfo {pages} {eadj4883} (\bibinfo {year}
  {2024})}\BibitemShut {NoStop}%
\bibitem [{\citenamefont {Krempask{\`y}}\ \emph {et~al.}(2024)\citenamefont {Krempask{\`y}}, \citenamefont {{\v{S}}mejkal}, \citenamefont {D’souza}, \citenamefont {Hajlaoui}, \citenamefont {Springholz}, \citenamefont {Uhl{\'\i}{\v{r}}ov{\'a}}, \citenamefont {Alarab}, \citenamefont {Constantinou}, \citenamefont {Strocov}, \citenamefont {Usanov} \emph {et~al.}}]{krempasky2024altermagnetic}%
  \BibitemOpen
  \bibfield  {author} {\bibinfo {author} {\bibfnamefont {J.}~\bibnamefont {Krempask{\`y}}}, \bibinfo {author} {\bibfnamefont {L.}~\bibnamefont {{\v{S}}mejkal}}, \bibinfo {author} {\bibfnamefont {S.}~\bibnamefont {D’souza}}, \bibinfo {author} {\bibfnamefont {M.}~\bibnamefont {Hajlaoui}}, \bibinfo {author} {\bibfnamefont {G.}~\bibnamefont {Springholz}}, \bibinfo {author} {\bibfnamefont {K.}~\bibnamefont {Uhl{\'\i}{\v{r}}ov{\'a}}}, \bibinfo {author} {\bibfnamefont {F.}~\bibnamefont {Alarab}}, \bibinfo {author} {\bibfnamefont {P.}~\bibnamefont {Constantinou}}, \bibinfo {author} {\bibfnamefont {V.}~\bibnamefont {Strocov}}, \bibinfo {author} {\bibfnamefont {D.}~\bibnamefont {Usanov}}, \emph {et~al.},\ }\bibfield  {title} {\bibinfo {title} {Altermagnetic lifting of kramers spin degeneracy},\ }\href {https://www.nature.com/articles/s41586-023-06907-7} {\bibfield  {journal} {\bibinfo  {journal} {Nature}\ }\textbf {\bibinfo {volume} {626}},\ \bibinfo {pages} {517} (\bibinfo {year} {2024})}\BibitemShut {NoStop}%
\bibitem [{\citenamefont {Zhu}\ \emph {et~al.}(2024)\citenamefont {Zhu}, \citenamefont {Chen}, \citenamefont {Liu}, \citenamefont {Liu}, \citenamefont {Liu}, \citenamefont {Zha}, \citenamefont {Qu}, \citenamefont {Hong}, \citenamefont {Li}, \citenamefont {Jiang} \emph {et~al.}}]{zhu2024observation}%
  \BibitemOpen
  \bibfield  {author} {\bibinfo {author} {\bibfnamefont {Y.-P.}\ \bibnamefont {Zhu}}, \bibinfo {author} {\bibfnamefont {X.}~\bibnamefont {Chen}}, \bibinfo {author} {\bibfnamefont {X.-R.}\ \bibnamefont {Liu}}, \bibinfo {author} {\bibfnamefont {Y.}~\bibnamefont {Liu}}, \bibinfo {author} {\bibfnamefont {P.}~\bibnamefont {Liu}}, \bibinfo {author} {\bibfnamefont {H.}~\bibnamefont {Zha}}, \bibinfo {author} {\bibfnamefont {G.}~\bibnamefont {Qu}}, \bibinfo {author} {\bibfnamefont {C.}~\bibnamefont {Hong}}, \bibinfo {author} {\bibfnamefont {J.}~\bibnamefont {Li}}, \bibinfo {author} {\bibfnamefont {Z.}~\bibnamefont {Jiang}}, \emph {et~al.},\ }\bibfield  {title} {\bibinfo {title} {Observation of plaid-like spin splitting in a noncoplanar antiferromagnet},\ }\href {https://www.nature.com/articles/s41586-024-07023-w} {\bibfield  {journal} {\bibinfo  {journal} {Nature}\ }\textbf {\bibinfo {volume} {626}},\ \bibinfo {pages} {523} (\bibinfo {year} {2024})}\BibitemShut {NoStop}%
\bibitem [{\citenamefont {Reimers}\ \emph {et~al.}(2024)\citenamefont {Reimers}, \citenamefont {Odenbreit}, \citenamefont {{\v{S}}mejkal}, \citenamefont {Strocov}, \citenamefont {Constantinou}, \citenamefont {Hellenes}, \citenamefont {Jaeschke~Ubiergo}, \citenamefont {Campos}, \citenamefont {Bharadwaj}, \citenamefont {Chakraborty} \emph {et~al.}}]{reimers2024direct}%
  \BibitemOpen
  \bibfield  {author} {\bibinfo {author} {\bibfnamefont {S.}~\bibnamefont {Reimers}}, \bibinfo {author} {\bibfnamefont {L.}~\bibnamefont {Odenbreit}}, \bibinfo {author} {\bibfnamefont {L.}~\bibnamefont {{\v{S}}mejkal}}, \bibinfo {author} {\bibfnamefont {V.~N.}\ \bibnamefont {Strocov}}, \bibinfo {author} {\bibfnamefont {P.}~\bibnamefont {Constantinou}}, \bibinfo {author} {\bibfnamefont {A.~B.}\ \bibnamefont {Hellenes}}, \bibinfo {author} {\bibfnamefont {R.}~\bibnamefont {Jaeschke~Ubiergo}}, \bibinfo {author} {\bibfnamefont {W.~H.}\ \bibnamefont {Campos}}, \bibinfo {author} {\bibfnamefont {V.~K.}\ \bibnamefont {Bharadwaj}}, \bibinfo {author} {\bibfnamefont {A.}~\bibnamefont {Chakraborty}}, \emph {et~al.},\ }\bibfield  {title} {\bibinfo {title} {Direct observation of altermagnetic band splitting in crsb thin films},\ }\href {https://www.nature.com/articles/s41467-024-46476-5} {\bibfield  {journal} {\bibinfo  {journal} {Nature Communications}\ }\textbf {\bibinfo {volume} {15}},\ \bibinfo {pages} {2116} (\bibinfo
  {year} {2024})}\BibitemShut {NoStop}%
\bibitem [{\citenamefont {Zhang}\ \emph {et~al.}(2025)\citenamefont {Zhang}, \citenamefont {Cheng}, \citenamefont {Yin}, \citenamefont {Liu}, \citenamefont {Deng}, \citenamefont {Qiao}, \citenamefont {Shi}, \citenamefont {Zhang}, \citenamefont {Lin}, \citenamefont {Liu} \emph {et~al.}}]{zhang2025crystal}%
  \BibitemOpen
  \bibfield  {author} {\bibinfo {author} {\bibfnamefont {F.}~\bibnamefont {Zhang}}, \bibinfo {author} {\bibfnamefont {X.}~\bibnamefont {Cheng}}, \bibinfo {author} {\bibfnamefont {Z.}~\bibnamefont {Yin}}, \bibinfo {author} {\bibfnamefont {C.}~\bibnamefont {Liu}}, \bibinfo {author} {\bibfnamefont {L.}~\bibnamefont {Deng}}, \bibinfo {author} {\bibfnamefont {Y.}~\bibnamefont {Qiao}}, \bibinfo {author} {\bibfnamefont {Z.}~\bibnamefont {Shi}}, \bibinfo {author} {\bibfnamefont {S.}~\bibnamefont {Zhang}}, \bibinfo {author} {\bibfnamefont {J.}~\bibnamefont {Lin}}, \bibinfo {author} {\bibfnamefont {Z.}~\bibnamefont {Liu}}, \emph {et~al.},\ }\bibfield  {title} {\bibinfo {title} {Crystal-symmetry-paired spin--valley locking in a layered room-temperature metallic altermagnet candidate},\ }\href {https://www.nature.com/articles/s41567-025-02864-2} {\bibfield  {journal} {\bibinfo  {journal} {Nature Physics}\ ,\ \bibinfo {pages} {1}} (\bibinfo {year} {2025})}\BibitemShut {NoStop}%
\bibitem [{\citenamefont {Gonzalez~Betancourt}\ \emph {et~al.}(2023)\citenamefont {Gonzalez~Betancourt}, \citenamefont {Zub{\'a}{\v{c}}}, \citenamefont {Gonzalez-Hernandez}, \citenamefont {Geishendorf}, \citenamefont {{\v{S}}ob{\'a}{\v{n}}}, \citenamefont {Springholz}, \citenamefont {Olejn{\'\i}k}, \citenamefont {{\v{S}}mejkal}, \citenamefont {Sinova}, \citenamefont {Jungwirth} \emph {et~al.}}]{gonzalez2023spontaneous}%
  \BibitemOpen
  \bibfield  {author} {\bibinfo {author} {\bibfnamefont {R.}~\bibnamefont {Gonzalez~Betancourt}}, \bibinfo {author} {\bibfnamefont {J.}~\bibnamefont {Zub{\'a}{\v{c}}}}, \bibinfo {author} {\bibfnamefont {R.}~\bibnamefont {Gonzalez-Hernandez}}, \bibinfo {author} {\bibfnamefont {K.}~\bibnamefont {Geishendorf}}, \bibinfo {author} {\bibfnamefont {Z.}~\bibnamefont {{\v{S}}ob{\'a}{\v{n}}}}, \bibinfo {author} {\bibfnamefont {G.}~\bibnamefont {Springholz}}, \bibinfo {author} {\bibfnamefont {K.}~\bibnamefont {Olejn{\'\i}k}}, \bibinfo {author} {\bibfnamefont {L.}~\bibnamefont {{\v{S}}mejkal}}, \bibinfo {author} {\bibfnamefont {J.}~\bibnamefont {Sinova}}, \bibinfo {author} {\bibfnamefont {T.}~\bibnamefont {Jungwirth}}, \emph {et~al.},\ }\bibfield  {title} {\bibinfo {title} {Spontaneous anomalous hall effect arising from an unconventional compensated magnetic phase in a semiconductor},\ }\href {https://link.aps.org/doi/10.1103/PhysRevLett.130.036702} {\bibfield  {journal} {\bibinfo  {journal} {Physical Review Letters}\
  }\textbf {\bibinfo {volume} {130}},\ \bibinfo {pages} {036702} (\bibinfo {year} {2023})}\BibitemShut {NoStop}%
\bibitem [{\citenamefont {Zhao}\ \emph {et~al.}(2025)\citenamefont {Zhao}, \citenamefont {Mao},\ and\ \citenamefont {Yan}}]{zhao2025nonlinear}%
  \BibitemOpen
  \bibfield  {author} {\bibinfo {author} {\bibfnamefont {Y.}~\bibnamefont {Zhao}}, \bibinfo {author} {\bibfnamefont {Z.}~\bibnamefont {Mao}},\ and\ \bibinfo {author} {\bibfnamefont {B.}~\bibnamefont {Yan}},\ }\bibfield  {title} {\bibinfo {title} {Nonlinear transport signatures of hidden symmetry breaking in a weyl altermagnet},\ }\href {https://doi.org/10.1103/f3vr-x4fx} {\bibfield  {journal} {\bibinfo  {journal} {Phys. Rev. B}\ }\textbf {\bibinfo {volume} {112}},\ \bibinfo {pages} {165127} (\bibinfo {year} {2025})}\BibitemShut {NoStop}%
\bibitem [{\citenamefont {Mali}\ \emph {et~al.}(2025)\citenamefont {Mali}, \citenamefont {Zhao}, \citenamefont {Wang}, \citenamefont {Sarker}, \citenamefont {Chen}, \citenamefont {Li}, \citenamefont {Zhu}, \citenamefont {Liu}, \citenamefont {Gopalan}, \citenamefont {Yan},\ and\ \citenamefont {Mao}}]{subin2025}%
  \BibitemOpen
  \bibfield  {author} {\bibinfo {author} {\bibfnamefont {S.}~\bibnamefont {Mali}}, \bibinfo {author} {\bibfnamefont {Y.}~\bibnamefont {Zhao}}, \bibinfo {author} {\bibfnamefont {Y.}~\bibnamefont {Wang}}, \bibinfo {author} {\bibfnamefont {S.}~\bibnamefont {Sarker}}, \bibinfo {author} {\bibfnamefont {Y.}~\bibnamefont {Chen}}, \bibinfo {author} {\bibfnamefont {Z.}~\bibnamefont {Li}}, \bibinfo {author} {\bibfnamefont {J.}~\bibnamefont {Zhu}}, \bibinfo {author} {\bibfnamefont {Y.}~\bibnamefont {Liu}}, \bibinfo {author} {\bibfnamefont {V.}~\bibnamefont {Gopalan}}, \bibinfo {author} {\bibfnamefont {B.}~\bibnamefont {Yan}},\ and\ \bibinfo {author} {\bibfnamefont {Z.}~\bibnamefont {Mao}},\ }\bibfield  {title} {\bibinfo {title} {Probing hidden symmetry and altermagnetism with sub-picometer sensitivity via nonlinear transport},\ }\href {https://arxiv.org/abs/2510.18144} {\bibfield  {journal} {\bibinfo  {journal} {arXiv:2510.18144}\ } (\bibinfo {year} {2025})}\BibitemShut {NoStop}%
\bibitem [{\citenamefont {Kluczyk}\ \emph {et~al.}(2024)\citenamefont {Kluczyk}, \citenamefont {Gas}, \citenamefont {Grzybowski}, \citenamefont {Skupinski}, \citenamefont {Borysiewicz}, \citenamefont {Fas}, \citenamefont {Suffczyski}, \citenamefont {Domagala}, \citenamefont {Grasza}, \citenamefont {Mycielski} \emph {et~al.}}]{kluczyk2024coexistence}%
  \BibitemOpen
  \bibfield  {author} {\bibinfo {author} {\bibfnamefont {K.}~\bibnamefont {Kluczyk}}, \bibinfo {author} {\bibfnamefont {K.}~\bibnamefont {Gas}}, \bibinfo {author} {\bibfnamefont {M.}~\bibnamefont {Grzybowski}}, \bibinfo {author} {\bibfnamefont {P.}~\bibnamefont {Skupinski}}, \bibinfo {author} {\bibfnamefont {M.}~\bibnamefont {Borysiewicz}}, \bibinfo {author} {\bibfnamefont {T.}~\bibnamefont {Fas}}, \bibinfo {author} {\bibfnamefont {J.}~\bibnamefont {Suffczyski}}, \bibinfo {author} {\bibfnamefont {J.}~\bibnamefont {Domagala}}, \bibinfo {author} {\bibfnamefont {K.}~\bibnamefont {Grasza}}, \bibinfo {author} {\bibfnamefont {A.}~\bibnamefont {Mycielski}}, \emph {et~al.},\ }\bibfield  {title} {\bibinfo {title} {Coexistence of anomalous hall effect and weak magnetization in a nominally collinear antiferromagnet mnte},\ }\href {https://link.aps.org/doi/10.1103/PhysRevB.110.155201} {\bibfield  {journal} {\bibinfo  {journal} {Physical Review B}\ }\textbf {\bibinfo {volume} {110}},\ \bibinfo {pages} {155201} (\bibinfo
  {year} {2024})}\BibitemShut {NoStop}%
\bibitem [{\citenamefont {Chilcote}\ \emph {et~al.}(2024)\citenamefont {Chilcote}, \citenamefont {Mazza}, \citenamefont {Lu}, \citenamefont {Gray}, \citenamefont {Tian}, \citenamefont {Deng}, \citenamefont {Moseley}, \citenamefont {Chen}, \citenamefont {Lapano}, \citenamefont {Gardner} \emph {et~al.}}]{chilcote2024stoichiometry}%
  \BibitemOpen
  \bibfield  {author} {\bibinfo {author} {\bibfnamefont {M.}~\bibnamefont {Chilcote}}, \bibinfo {author} {\bibfnamefont {A.~R.}\ \bibnamefont {Mazza}}, \bibinfo {author} {\bibfnamefont {Q.}~\bibnamefont {Lu}}, \bibinfo {author} {\bibfnamefont {I.}~\bibnamefont {Gray}}, \bibinfo {author} {\bibfnamefont {Q.}~\bibnamefont {Tian}}, \bibinfo {author} {\bibfnamefont {Q.}~\bibnamefont {Deng}}, \bibinfo {author} {\bibfnamefont {D.}~\bibnamefont {Moseley}}, \bibinfo {author} {\bibfnamefont {A.-H.}\ \bibnamefont {Chen}}, \bibinfo {author} {\bibfnamefont {J.}~\bibnamefont {Lapano}}, \bibinfo {author} {\bibfnamefont {J.~S.}\ \bibnamefont {Gardner}}, \emph {et~al.},\ }\bibfield  {title} {\bibinfo {title} {Stoichiometry-induced ferromagnetism in altermagnetic candidate mnte},\ }\href {https://advanced.onlinelibrary.wiley.com/doi/10.1002/adfm.202405829} {\bibfield  {journal} {\bibinfo  {journal} {Advanced Functional Materials}\ }\textbf {\bibinfo {volume} {34}},\ \bibinfo {pages} {2405829} (\bibinfo {year}
  {2024})}\BibitemShut {NoStop}%
\bibitem [{\citenamefont {Bey}\ \emph {et~al.}(2024)\citenamefont {Bey}, \citenamefont {Fields}, \citenamefont {Combs}, \citenamefont {Markus}, \citenamefont {Beke}, \citenamefont {Wang}, \citenamefont {Ievlev}, \citenamefont {Zhukovskyi}, \citenamefont {Orlova}, \citenamefont {Forro} \emph {et~al.}}]{bey2024unexpected}%
  \BibitemOpen
  \bibfield  {author} {\bibinfo {author} {\bibfnamefont {S.}~\bibnamefont {Bey}}, \bibinfo {author} {\bibfnamefont {S.~S.}\ \bibnamefont {Fields}}, \bibinfo {author} {\bibfnamefont {N.~G.}\ \bibnamefont {Combs}}, \bibinfo {author} {\bibfnamefont {B.~G.}\ \bibnamefont {Markus}}, \bibinfo {author} {\bibfnamefont {D.}~\bibnamefont {Beke}}, \bibinfo {author} {\bibfnamefont {J.}~\bibnamefont {Wang}}, \bibinfo {author} {\bibfnamefont {A.~V.}\ \bibnamefont {Ievlev}}, \bibinfo {author} {\bibfnamefont {M.}~\bibnamefont {Zhukovskyi}}, \bibinfo {author} {\bibfnamefont {T.}~\bibnamefont {Orlova}}, \bibinfo {author} {\bibfnamefont {L.}~\bibnamefont {Forro}}, \emph {et~al.},\ }\bibfield  {title} {\bibinfo {title} {Unexpected tuning of the anomalous hall effect in altermagnetic mnte thin films},\ }\href {https://arxiv.org/abs/2409.04567} {\bibfield  {journal} {\bibinfo  {journal} {arXiv preprint arXiv:2409.04567}\ } (\bibinfo {year} {2024})}\BibitemShut {NoStop}%
\bibitem [{\citenamefont {Liu}\ \emph {et~al.}(2025{\natexlab{b}})\citenamefont {Liu}, \citenamefont {Xu}, \citenamefont {DeStefano}, \citenamefont {Rosenberg}, \citenamefont {Zhang}, \citenamefont {Li}, \citenamefont {Stone}, \citenamefont {Ye}, \citenamefont {Cong}, \citenamefont {Pan} \emph {et~al.}}]{liu2025strain}%
  \BibitemOpen
  \bibfield  {author} {\bibinfo {author} {\bibfnamefont {Z.}~\bibnamefont {Liu}}, \bibinfo {author} {\bibfnamefont {S.}~\bibnamefont {Xu}}, \bibinfo {author} {\bibfnamefont {J.~M.}\ \bibnamefont {DeStefano}}, \bibinfo {author} {\bibfnamefont {E.}~\bibnamefont {Rosenberg}}, \bibinfo {author} {\bibfnamefont {T.}~\bibnamefont {Zhang}}, \bibinfo {author} {\bibfnamefont {J.}~\bibnamefont {Li}}, \bibinfo {author} {\bibfnamefont {M.~B.}\ \bibnamefont {Stone}}, \bibinfo {author} {\bibfnamefont {F.}~\bibnamefont {Ye}}, \bibinfo {author} {\bibfnamefont {R.}~\bibnamefont {Cong}}, \bibinfo {author} {\bibfnamefont {S.}~\bibnamefont {Pan}}, \emph {et~al.},\ }\bibfield  {title} {\bibinfo {title} {Strain-tunable anomalous hall effect in hexagonal mnte},\ }\href {https://arxiv.org/abs/2509.19582} {\bibfield  {journal} {\bibinfo  {journal} {arXiv preprint arXiv:2509.19582}\ } (\bibinfo {year} {2025}{\natexlab{b}})}\BibitemShut {NoStop}%
\bibitem [{\citenamefont {Smolenski}\ \emph {et~al.}(2025)\citenamefont {Smolenski}, \citenamefont {Mao}, \citenamefont {Zhang}, \citenamefont {Guo}, \citenamefont {Shawon}, \citenamefont {Xu}, \citenamefont {Downey}, \citenamefont {Musall}, \citenamefont {Yi}, \citenamefont {Xie} \emph {et~al.}}]{smolenski2025strain}%
  \BibitemOpen
  \bibfield  {author} {\bibinfo {author} {\bibfnamefont {S.}~\bibnamefont {Smolenski}}, \bibinfo {author} {\bibfnamefont {N.}~\bibnamefont {Mao}}, \bibinfo {author} {\bibfnamefont {D.}~\bibnamefont {Zhang}}, \bibinfo {author} {\bibfnamefont {Y.}~\bibnamefont {Guo}}, \bibinfo {author} {\bibfnamefont {A.}~\bibnamefont {Shawon}}, \bibinfo {author} {\bibfnamefont {M.}~\bibnamefont {Xu}}, \bibinfo {author} {\bibfnamefont {E.}~\bibnamefont {Downey}}, \bibinfo {author} {\bibfnamefont {T.}~\bibnamefont {Musall}}, \bibinfo {author} {\bibfnamefont {M.}~\bibnamefont {Yi}}, \bibinfo {author} {\bibfnamefont {W.}~\bibnamefont {Xie}}, \emph {et~al.},\ }\bibfield  {title} {\bibinfo {title} {Strain-tunability of the multipolar berry curvature in altermagnet mnte},\ }\href {https://arxiv.org/abs/2509.21481} {\bibfield  {journal} {\bibinfo  {journal} {arXiv preprint arXiv:2509.21481}\ } (\bibinfo {year} {2025})}\BibitemShut {NoStop}%
\bibitem [{\citenamefont {Zhou}\ \emph {et~al.}()\citenamefont {Zhou}, \citenamefont {Yan}, \citenamefont {Rong}, \citenamefont {Zhao}, \citenamefont {Xiao}, \citenamefont {Lai}, \citenamefont {Xi}, \citenamefont {Wang}, \citenamefont {Adhikari}, \citenamefont {Tiwari}, \citenamefont {Lin}, \citenamefont {Manue}, \citenamefont {Orlandi}, \citenamefont {Khalyavin}, \citenamefont {Grutter}, \citenamefont {Liu}, \citenamefont {Yan},\ and\ \citenamefont {Chang}}]{zhou2026}%
  \BibitemOpen
  \bibfield  {author} {\bibinfo {author} {\bibfnamefont {L.-J.}\ \bibnamefont {Zhou}}, \bibinfo {author} {\bibfnamefont {Z.-J.}\ \bibnamefont {Yan}}, \bibinfo {author} {\bibfnamefont {H.}~\bibnamefont {Rong}}, \bibinfo {author} {\bibfnamefont {Y.}~\bibnamefont {Zhao}}, \bibinfo {author} {\bibfnamefont {P.}~\bibnamefont {Xiao}}, \bibinfo {author} {\bibfnamefont {L.-K.}\ \bibnamefont {Lai}}, \bibinfo {author} {\bibfnamefont {Z.}~\bibnamefont {Xi}}, \bibinfo {author} {\bibfnamefont {K.}~\bibnamefont {Wang}}, \bibinfo {author} {\bibfnamefont {T.}~\bibnamefont {Adhikari}}, \bibinfo {author} {\bibfnamefont {G.~P.}\ \bibnamefont {Tiwari}}, \bibinfo {author} {\bibfnamefont {Z.}~\bibnamefont {Lin}}, \bibinfo {author} {\bibfnamefont {P.}~\bibnamefont {Manue}}, \bibinfo {author} {\bibfnamefont {F.}~\bibnamefont {Orlandi}}, \bibinfo {author} {\bibfnamefont {D.}~\bibnamefont {Khalyavin}}, \bibinfo {author} {\bibfnamefont {A.~J.}\ \bibnamefont {Grutter}}, \bibinfo {author} {\bibfnamefont {C.-X.}\ \bibnamefont {Liu}},
  \bibinfo {author} {\bibfnamefont {B.}~\bibnamefont {Yan}},\ and\ \bibinfo {author} {\bibfnamefont {C.-Z.}\ \bibnamefont {Chang}},\ }\bibfield  {title} {\bibinfo {title} {Surface-state-driven anomalous hall effect in altermagnetic mnte films},\ }\href {https://arxiv.org/abs/2602.09363} {\bibinfo  {journal} {arXiv:2602.09363}\ }\BibitemShut {NoStop}%
\bibitem [{\citenamefont {Osumi}\ \emph {et~al.}(2024)\citenamefont {Osumi}, \citenamefont {Souma}, \citenamefont {Aoyama}, \citenamefont {Yamauchi}, \citenamefont {Honma}, \citenamefont {Nakayama}, \citenamefont {Takahashi}, \citenamefont {Ohgushi},\ and\ \citenamefont {Sato}}]{Osumi2024prb}%
  \BibitemOpen
\bibfield  {journal} {  }\bibfield  {author} {\bibinfo {author} {\bibfnamefont {T.}~\bibnamefont {Osumi}}, \bibinfo {author} {\bibfnamefont {S.}~\bibnamefont {Souma}}, \bibinfo {author} {\bibfnamefont {T.}~\bibnamefont {Aoyama}}, \bibinfo {author} {\bibfnamefont {K.}~\bibnamefont {Yamauchi}}, \bibinfo {author} {\bibfnamefont {A.}~\bibnamefont {Honma}}, \bibinfo {author} {\bibfnamefont {K.}~\bibnamefont {Nakayama}}, \bibinfo {author} {\bibfnamefont {T.}~\bibnamefont {Takahashi}}, \bibinfo {author} {\bibfnamefont {K.}~\bibnamefont {Ohgushi}},\ and\ \bibinfo {author} {\bibfnamefont {T.}~\bibnamefont {Sato}},\ }\bibfield  {title} {\bibinfo {title} {Observation of a giant band splitting in altermagnetic mnte},\ }\href {https://doi.org/10.1103/PhysRevB.109.115102} {\bibfield  {journal} {\bibinfo  {journal} {Phys. Rev. B}\ }\textbf {\bibinfo {volume} {109}},\ \bibinfo {pages} {115102} (\bibinfo {year} {2024})}\BibitemShut {NoStop}%
\bibitem [{\citenamefont {Hajlaoui}\ \emph {et~al.}(2024)\citenamefont {Hajlaoui}, \citenamefont {Wilfred~D'Souza}, \citenamefont {{\v{S}}mejkal}, \citenamefont {Kriegner}, \citenamefont {Krizman}, \citenamefont {Zakusylo}, \citenamefont {Olszowska}, \citenamefont {Caha}, \citenamefont {Michali{\v{c}}ka}, \citenamefont {S{\'a}nchez-Barriga} \emph {et~al.}}]{hajlaoui2024temperature}%
  \BibitemOpen
  \bibfield  {author} {\bibinfo {author} {\bibfnamefont {M.}~\bibnamefont {Hajlaoui}}, \bibinfo {author} {\bibfnamefont {S.}~\bibnamefont {Wilfred~D'Souza}}, \bibinfo {author} {\bibfnamefont {L.}~\bibnamefont {{\v{S}}mejkal}}, \bibinfo {author} {\bibfnamefont {D.}~\bibnamefont {Kriegner}}, \bibinfo {author} {\bibfnamefont {G.}~\bibnamefont {Krizman}}, \bibinfo {author} {\bibfnamefont {T.}~\bibnamefont {Zakusylo}}, \bibinfo {author} {\bibfnamefont {N.}~\bibnamefont {Olszowska}}, \bibinfo {author} {\bibfnamefont {O.}~\bibnamefont {Caha}}, \bibinfo {author} {\bibfnamefont {J.}~\bibnamefont {Michali{\v{c}}ka}}, \bibinfo {author} {\bibfnamefont {J.}~\bibnamefont {S{\'a}nchez-Barriga}}, \emph {et~al.},\ }\bibfield  {title} {\bibinfo {title} {Temperature dependence of relativistic valence band splitting induced by an altermagnetic phase transition},\ }\href {https://advanced.onlinelibrary.wiley.com/doi/full/10.1002/adma.202314076} {\bibfield  {journal} {\bibinfo  {journal} {Advanced Materials}\ }\textbf {\bibinfo
  {volume} {36}},\ \bibinfo {pages} {2314076} (\bibinfo {year} {2024})}\BibitemShut {NoStop}%
\bibitem [{\citenamefont {Kresse}\ and\ \citenamefont {Furthm{\"u}ller}(1996)}]{kresse1996efficient}%
  \BibitemOpen
  \bibfield  {author} {\bibinfo {author} {\bibfnamefont {G.}~\bibnamefont {Kresse}}\ and\ \bibinfo {author} {\bibfnamefont {J.}~\bibnamefont {Furthm{\"u}ller}},\ }\bibfield  {title} {\bibinfo {title} {Efficient iterative schemes for ab initio total-energy calculations using a plane-wave basis set},\ }\href {https://journals.aps.org/prb/abstract/10.1103/PhysRevB.54.11169} {\bibfield  {journal} {\bibinfo  {journal} {Physical review B}\ }\textbf {\bibinfo {volume} {54}},\ \bibinfo {pages} {11169} (\bibinfo {year} {1996})}\BibitemShut {NoStop}%
\bibitem [{\citenamefont {Kresse}\ and\ \citenamefont {Joubert}(1999)}]{kresse1999ultrasoft}%
  \BibitemOpen
  \bibfield  {author} {\bibinfo {author} {\bibfnamefont {G.}~\bibnamefont {Kresse}}\ and\ \bibinfo {author} {\bibfnamefont {D.}~\bibnamefont {Joubert}},\ }\bibfield  {title} {\bibinfo {title} {From ultrasoft pseudopotentials to the projector augmented-wave method},\ }\href {https://journals.aps.org/prb/abstract/10.1103/PhysRevB.59.1758} {\bibfield  {journal} {\bibinfo  {journal} {Physical review b}\ }\textbf {\bibinfo {volume} {59}},\ \bibinfo {pages} {1758} (\bibinfo {year} {1999})}\BibitemShut {NoStop}%
\bibitem [{\citenamefont {Perdew}\ \emph {et~al.}(1996)\citenamefont {Perdew}, \citenamefont {Burke},\ and\ \citenamefont {Ernzerhof}}]{perdew1996generalized}%
  \BibitemOpen
  \bibfield  {author} {\bibinfo {author} {\bibfnamefont {J.~P.}\ \bibnamefont {Perdew}}, \bibinfo {author} {\bibfnamefont {K.}~\bibnamefont {Burke}},\ and\ \bibinfo {author} {\bibfnamefont {M.}~\bibnamefont {Ernzerhof}},\ }\bibfield  {title} {\bibinfo {title} {Generalized gradient approximation made simple},\ }\href@noop {} {\bibfield  {journal} {\bibinfo  {journal} {Physical review letters}\ }\textbf {\bibinfo {volume} {77}},\ \bibinfo {pages} {3865} (\bibinfo {year} {1996})}\BibitemShut {NoStop}%
\bibitem [{\citenamefont {Dudarev}\ \emph {et~al.}(1998)\citenamefont {Dudarev}, \citenamefont {Botton}, \citenamefont {Savrasov}, \citenamefont {Humphreys},\ and\ \citenamefont {Sutton}}]{dudarev1998electron}%
  \BibitemOpen
  \bibfield  {author} {\bibinfo {author} {\bibfnamefont {S.~L.}\ \bibnamefont {Dudarev}}, \bibinfo {author} {\bibfnamefont {G.~A.}\ \bibnamefont {Botton}}, \bibinfo {author} {\bibfnamefont {S.~Y.}\ \bibnamefont {Savrasov}}, \bibinfo {author} {\bibfnamefont {C.}~\bibnamefont {Humphreys}},\ and\ \bibinfo {author} {\bibfnamefont {A.~P.}\ \bibnamefont {Sutton}},\ }\bibfield  {title} {\bibinfo {title} {Electron-energy-loss spectra and the structural stability of nickel oxide: An lsda+ u study},\ }\href {https://journals.aps.org/prb/abstract/10.1103/PhysRevB.57.1505} {\bibfield  {journal} {\bibinfo  {journal} {Physical Review B}\ }\textbf {\bibinfo {volume} {57}},\ \bibinfo {pages} {1505} (\bibinfo {year} {1998})}\BibitemShut {NoStop}%
\bibitem [{\citenamefont {Souza}\ \emph {et~al.}(2001)\citenamefont {Souza}, \citenamefont {Marzari},\ and\ \citenamefont {Vanderbilt}}]{souza2001maximally}%
  \BibitemOpen
  \bibfield  {author} {\bibinfo {author} {\bibfnamefont {I.}~\bibnamefont {Souza}}, \bibinfo {author} {\bibfnamefont {N.}~\bibnamefont {Marzari}},\ and\ \bibinfo {author} {\bibfnamefont {D.}~\bibnamefont {Vanderbilt}},\ }\bibfield  {title} {\bibinfo {title} {Maximally localized wannier functions for entangled energy bands},\ }\href {https://journals.aps.org/prb/abstract/10.1103/PhysRevB.65.035109} {\bibfield  {journal} {\bibinfo  {journal} {Physical Review B}\ }\textbf {\bibinfo {volume} {65}},\ \bibinfo {pages} {035109} (\bibinfo {year} {2001})}\BibitemShut {NoStop}%
\bibitem [{\citenamefont {Varnava}\ and\ \citenamefont {Vanderbilt}(2018)}]{Varnava2018}%
  \BibitemOpen
  \bibfield  {author} {\bibinfo {author} {\bibfnamefont {N.}~\bibnamefont {Varnava}}\ and\ \bibinfo {author} {\bibfnamefont {D.}~\bibnamefont {Vanderbilt}},\ }\bibfield  {title} {\bibinfo {title} {Surfaces of axion insulators},\ }\href {https://doi.org/10.1103/PhysRevB.98.245117} {\bibfield  {journal} {\bibinfo  {journal} {Phys. Rev. B}\ }\textbf {\bibinfo {volume} {98}},\ \bibinfo {pages} {245117} (\bibinfo {year} {2018})}\BibitemShut {NoStop}%
\bibitem [{\citenamefont {Zhao}\ \emph {et~al.}(2024)\citenamefont {Zhao}, \citenamefont {Jiang}, \citenamefont {Bae}, \citenamefont {Das}, \citenamefont {Li}, \citenamefont {Liu},\ and\ \citenamefont {Yan}}]{Zhao2024}%
  \BibitemOpen
  \bibfield  {author} {\bibinfo {author} {\bibfnamefont {Y.}~\bibnamefont {Zhao}}, \bibinfo {author} {\bibfnamefont {Y.}~\bibnamefont {Jiang}}, \bibinfo {author} {\bibfnamefont {H.}~\bibnamefont {Bae}}, \bibinfo {author} {\bibfnamefont {K.}~\bibnamefont {Das}}, \bibinfo {author} {\bibfnamefont {Y.}~\bibnamefont {Li}}, \bibinfo {author} {\bibfnamefont {C.-X.}\ \bibnamefont {Liu}},\ and\ \bibinfo {author} {\bibfnamefont {B.}~\bibnamefont {Yan}},\ }\bibfield  {title} {\bibinfo {title} {Hybrid-order topology in unconventional magnets of eu-based zintl compounds with surface-dependent quantum geometry},\ }\href {https://doi.org/10.1103/PhysRevB.110.205111} {\bibfield  {journal} {\bibinfo  {journal} {Phys. Rev. B}\ }\textbf {\bibinfo {volume} {110}},\ \bibinfo {pages} {205111} (\bibinfo {year} {2024})}\BibitemShut {NoStop}%
\bibitem [{\citenamefont {Litvin}(1977)}]{litvin1977spin}%
  \BibitemOpen
  \bibfield  {author} {\bibinfo {author} {\bibfnamefont {D.~B.}\ \bibnamefont {Litvin}},\ }\bibfield  {title} {\bibinfo {title} {Spin point groups},\ }\href@noop {} {\bibfield  {journal} {\bibinfo  {journal} {Foundations of Crystallography}\ }\textbf {\bibinfo {volume} {33}},\ \bibinfo {pages} {279} (\bibinfo {year} {1977})}\BibitemShut {NoStop}%
\bibitem [{\citenamefont {Chen}\ \emph {et~al.}(2024)\citenamefont {Chen}, \citenamefont {Ren}, \citenamefont {Zhu}, \citenamefont {Yu}, \citenamefont {Zhang}, \citenamefont {Liu}, \citenamefont {Li}, \citenamefont {Liu}, \citenamefont {Li},\ and\ \citenamefont {Liu}}]{chen2024enumeration}%
  \BibitemOpen
  \bibfield  {author} {\bibinfo {author} {\bibfnamefont {X.}~\bibnamefont {Chen}}, \bibinfo {author} {\bibfnamefont {J.}~\bibnamefont {Ren}}, \bibinfo {author} {\bibfnamefont {Y.}~\bibnamefont {Zhu}}, \bibinfo {author} {\bibfnamefont {Y.}~\bibnamefont {Yu}}, \bibinfo {author} {\bibfnamefont {A.}~\bibnamefont {Zhang}}, \bibinfo {author} {\bibfnamefont {P.}~\bibnamefont {Liu}}, \bibinfo {author} {\bibfnamefont {J.}~\bibnamefont {Li}}, \bibinfo {author} {\bibfnamefont {Y.}~\bibnamefont {Liu}}, \bibinfo {author} {\bibfnamefont {C.}~\bibnamefont {Li}},\ and\ \bibinfo {author} {\bibfnamefont {Q.}~\bibnamefont {Liu}},\ }\bibfield  {title} {\bibinfo {title} {Enumeration and representation theory of spin space groups},\ }\href@noop {} {\bibfield  {journal} {\bibinfo  {journal} {Physical Review X}\ }\textbf {\bibinfo {volume} {14}},\ \bibinfo {pages} {031038} (\bibinfo {year} {2024})}\BibitemShut {NoStop}%
\bibitem [{\citenamefont {Jiang}\ \emph {et~al.}(2024)\citenamefont {Jiang}, \citenamefont {Song}, \citenamefont {Zhu}, \citenamefont {Fang}, \citenamefont {Weng}, \citenamefont {Liu}, \citenamefont {Yang},\ and\ \citenamefont {Fang}}]{jiang2024enumeration}%
  \BibitemOpen
  \bibfield  {author} {\bibinfo {author} {\bibfnamefont {Y.}~\bibnamefont {Jiang}}, \bibinfo {author} {\bibfnamefont {Z.}~\bibnamefont {Song}}, \bibinfo {author} {\bibfnamefont {T.}~\bibnamefont {Zhu}}, \bibinfo {author} {\bibfnamefont {Z.}~\bibnamefont {Fang}}, \bibinfo {author} {\bibfnamefont {H.}~\bibnamefont {Weng}}, \bibinfo {author} {\bibfnamefont {Z.-X.}\ \bibnamefont {Liu}}, \bibinfo {author} {\bibfnamefont {J.}~\bibnamefont {Yang}},\ and\ \bibinfo {author} {\bibfnamefont {C.}~\bibnamefont {Fang}},\ }\bibfield  {title} {\bibinfo {title} {Enumeration of spin-space groups: Toward a complete description of symmetries of magnetic orders},\ }\href@noop {} {\bibfield  {journal} {\bibinfo  {journal} {Physical Review X}\ }\textbf {\bibinfo {volume} {14}},\ \bibinfo {pages} {031039} (\bibinfo {year} {2024})}\BibitemShut {NoStop}%
\bibitem [{\citenamefont {Xiao}\ \emph {et~al.}(2024)\citenamefont {Xiao}, \citenamefont {Zhao}, \citenamefont {Li}, \citenamefont {Shindou},\ and\ \citenamefont {Song}}]{xiao2024spin}%
  \BibitemOpen
  \bibfield  {author} {\bibinfo {author} {\bibfnamefont {Z.}~\bibnamefont {Xiao}}, \bibinfo {author} {\bibfnamefont {J.}~\bibnamefont {Zhao}}, \bibinfo {author} {\bibfnamefont {Y.}~\bibnamefont {Li}}, \bibinfo {author} {\bibfnamefont {R.}~\bibnamefont {Shindou}},\ and\ \bibinfo {author} {\bibfnamefont {Z.-D.}\ \bibnamefont {Song}},\ }\bibfield  {title} {\bibinfo {title} {Spin space groups: Full classification and applications},\ }\href@noop {} {\bibfield  {journal} {\bibinfo  {journal} {Physical Review X}\ }\textbf {\bibinfo {volume} {14}},\ \bibinfo {pages} {031037} (\bibinfo {year} {2024})}\BibitemShut {NoStop}%
\bibitem [{\citenamefont {Das}\ \emph {et~al.}(2025)\citenamefont {Das}, \citenamefont {Zhao},\ and\ \citenamefont {Yan}}]{das2025surface}%
  \BibitemOpen
  \bibfield  {author} {\bibinfo {author} {\bibfnamefont {K.}~\bibnamefont {Das}}, \bibinfo {author} {\bibfnamefont {Y.}~\bibnamefont {Zhao}},\ and\ \bibinfo {author} {\bibfnamefont {B.}~\bibnamefont {Yan}},\ }\bibfield  {title} {\bibinfo {title} {Surface-dominated quantum-metric-induced nonlinear transport in the layered antiferromagnet crsbr},\ }\href@noop {} {\bibfield  {journal} {\bibinfo  {journal} {Nano letters}\ }\textbf {\bibinfo {volume} {25}},\ \bibinfo {pages} {9189} (\bibinfo {year} {2025})}\BibitemShut {NoStop}%
\bibitem [{sup()}]{supplement}%
  \BibitemOpen
  \bibfield  {title} {\bibinfo {title} {Supplementary materials, including model hamiltonian and additional calculation results},\ }\href@noop {} {\bibinfo  {journal} {arXiv}\ }\BibitemShut {NoStop}%
\bibitem [{\citenamefont {Li}\ \emph {et~al.}(2019)\citenamefont {Li}, \citenamefont {Li}, \citenamefont {Du}, \citenamefont {Wang}, \citenamefont {Gu}, \citenamefont {Zhang}, \citenamefont {He}, \citenamefont {Duan},\ and\ \citenamefont {Xu}}]{li2019intrinsic}%
  \BibitemOpen
\bibfield  {journal} {  }\bibfield  {author} {\bibinfo {author} {\bibfnamefont {J.}~\bibnamefont {Li}}, \bibinfo {author} {\bibfnamefont {Y.}~\bibnamefont {Li}}, \bibinfo {author} {\bibfnamefont {S.}~\bibnamefont {Du}}, \bibinfo {author} {\bibfnamefont {Z.}~\bibnamefont {Wang}}, \bibinfo {author} {\bibfnamefont {B.-L.}\ \bibnamefont {Gu}}, \bibinfo {author} {\bibfnamefont {S.-C.}\ \bibnamefont {Zhang}}, \bibinfo {author} {\bibfnamefont {K.}~\bibnamefont {He}}, \bibinfo {author} {\bibfnamefont {W.}~\bibnamefont {Duan}},\ and\ \bibinfo {author} {\bibfnamefont {Y.}~\bibnamefont {Xu}},\ }\bibfield  {title} {\bibinfo {title} {Intrinsic magnetic topological insulators in van der waals layered mnbi2te4-family materials},\ }\href {https://www.science.org/doi/epdf/10.1126/sciadv.aaw5685} {\bibfield  {journal} {\bibinfo  {journal} {Science advances}\ }\textbf {\bibinfo {volume} {5}},\ \bibinfo {pages} {eaaw5685} (\bibinfo {year} {2019})}\BibitemShut {NoStop}%
\bibitem [{\citenamefont {Sun}\ \emph {et~al.}(2009)\citenamefont {Sun}, \citenamefont {Yao}, \citenamefont {Fradkin},\ and\ \citenamefont {Kivelson}}]{sun2009topological}%
  \BibitemOpen
  \bibfield  {author} {\bibinfo {author} {\bibfnamefont {K.}~\bibnamefont {Sun}}, \bibinfo {author} {\bibfnamefont {H.}~\bibnamefont {Yao}}, \bibinfo {author} {\bibfnamefont {E.}~\bibnamefont {Fradkin}},\ and\ \bibinfo {author} {\bibfnamefont {S.~A.}\ \bibnamefont {Kivelson}},\ }\bibfield  {title} {\bibinfo {title} {Topological insulators and nematic phases from spontaneous symmetry breaking<? format?> in 2d fermi systems with a quadratic band crossing},\ }\href@noop {} {\bibfield  {journal} {\bibinfo  {journal} {Physical review letters}\ }\textbf {\bibinfo {volume} {103}},\ \bibinfo {pages} {046811} (\bibinfo {year} {2009})}\BibitemShut {NoStop}%
\bibitem [{\citenamefont {Tasker}(1979)}]{tasker1979stability}%
  \BibitemOpen
  \bibfield  {author} {\bibinfo {author} {\bibfnamefont {P.}~\bibnamefont {Tasker}},\ }\bibfield  {title} {\bibinfo {title} {The stability of ionic crystal surfaces},\ }\href {https://iopscience.iop.org/article/10.1088/0022-3719/12/22/036} {\bibfield  {journal} {\bibinfo  {journal} {Journal of Physics C: Solid State Physics}\ }\textbf {\bibinfo {volume} {12}},\ \bibinfo {pages} {4977} (\bibinfo {year} {1979})}\BibitemShut {NoStop}%
\bibitem [{\citenamefont {Noguera}(2000)}]{noguera2000polar}%
  \BibitemOpen
  \bibfield  {author} {\bibinfo {author} {\bibfnamefont {C.}~\bibnamefont {Noguera}},\ }\bibfield  {title} {\bibinfo {title} {Polar oxide surfaces},\ }\href {https://iopscience.iop.org/article/10.1088/0953-8984/12/31/201} {\bibfield  {journal} {\bibinfo  {journal} {Journal of Physics: Condensed Matter}\ }\textbf {\bibinfo {volume} {12}},\ \bibinfo {pages} {R367} (\bibinfo {year} {2000})}\BibitemShut {NoStop}%
\bibitem [{\citenamefont {Nagaosa}\ \emph {et~al.}(2010)\citenamefont {Nagaosa}, \citenamefont {Sinova}, \citenamefont {Onoda}, \citenamefont {MacDonald},\ and\ \citenamefont {Ong}}]{RevModPhys.82.1539}%
  \BibitemOpen
  \bibfield  {author} {\bibinfo {author} {\bibfnamefont {N.}~\bibnamefont {Nagaosa}}, \bibinfo {author} {\bibfnamefont {J.}~\bibnamefont {Sinova}}, \bibinfo {author} {\bibfnamefont {S.}~\bibnamefont {Onoda}}, \bibinfo {author} {\bibfnamefont {A.~H.}\ \bibnamefont {MacDonald}},\ and\ \bibinfo {author} {\bibfnamefont {N.~P.}\ \bibnamefont {Ong}},\ }\bibfield  {title} {\bibinfo {title} {Anomalous hall effect},\ }\href {https://doi.org/10.1103/RevModPhys.82.1539} {\bibfield  {journal} {\bibinfo  {journal} {Rev. Mod. Phys.}\ }\textbf {\bibinfo {volume} {82}},\ \bibinfo {pages} {1539} (\bibinfo {year} {2010})}\BibitemShut {NoStop}%
\end{thebibliography}

%

\end{document}